\documentclass[aps,12pt,
		prd,
		reprint,
		onecolumn,
		superscriptaddress,
		shortbibliography,
		nofootinbib,
		floatfix,
		notitlepage
		]{revtex4-1}

\usepackage{amsmath,amssymb,amsfonts}

\usepackage{natbib}
\usepackage[T1]{fontenc}
\usepackage[latin1]{inputenc}

\usepackage[dvipsnames]{xcolor}

\usepackage{hyperref}
\hypersetup{colorlinks=true,citecolor=Cyan,urlcolor=Red}

\usepackage{mathrsfs}
\usepackage{bbm}
\usepackage{slashed}
\DeclareSymbolFont{rsfs}{U}{rsfs}{m}{n}
\DeclareSymbolFontAlphabet{\mathrsfs}{rsfs}
\usepackage[mathscr]{eucal}	
\usepackage[normalem]{ulem}
\usepackage{verbatim}
\usepackage{bm}

\usepackage{graphicx}

\definecolor{bc}{rgb}{0, 0.7, 0.0}

\newcommand{\be}{\begin{equation}}
\newcommand{\ee}{\end{equation}}
\newcommand{\bi}{\begin{itemize}}
\newcommand{\ei}{\end{itemize}}
\newcommand{\bea}{\begin{eqnarray}}
\newcommand{\eea}{\end{eqnarray}}
\newcommand{\ud}{\mathrm{d}}		

\newcommand{\Ai}{{\text{Ai}}}

\newcommand{\LCm}{{\scriptscriptstyle -}} 
\newcommand{\LCp}{{\scriptscriptstyle +}}
\newcommand{\LCpm}{{\scriptscriptstyle \pm}}

\newcommand{\LCperp}{{\scriptscriptstyle \perp}}

\newcommand{\lf}{\text{\tiny LF}}

\begin{document}
%


\title{Absorption cross section in an intense plane wave background}

\author{A. Ilderton}
\affiliation{Centre for Mathematical Sciences, University of Plymouth, Plymouth, PL4 8AA, UK}
\author{B. King}
\affiliation{Centre for Mathematical Sciences, University of Plymouth, Plymouth, PL4 8AA, UK}
\author{A. J. MacLeod}
\email{alexander.macleod@plymouth.ac.uk}
\affiliation{Centre for Mathematical Sciences, University of Plymouth, Plymouth, PL4 8AA, UK}

\begin{abstract}	
We consider the absorption of probe photons by electrons in the presence of an intense, pulsed, {background} field. Our analysis reveals {an interplay between regularisation and gauge invariance which distinguishes absorption from its crossing-symmetric processes, as well as a physical interpretation of absorption in terms of degenerate processes in the weak field limit. In the strong field limit} we develop a locally constant field approximation (LCFA) for absorption which also exhibits new features. We benchmark the LCFA against exact analytical calculations and explore its regime of validity. Pulse shape effects are also investigated, as well as infra-red and collinear limits of the absorption process. 
\end{abstract}
\pacs{}
\maketitle

\section{Introduction}
A number of upcoming experiments  aim to probe quantum processes in intense laser fields, including pair production~\cite{Abramowicz:2019gvx}, vacuum birefringence~\cite{King:2015tba}, and conjectured regimes where all current perturbative approaches to QED break {down~\cite{Fedotov:2016afw,Yakimenko:2018kih,Blackburn:2018tsn,Baumann:2018ovl}}.  However, a series of theoretical investigations~\cite{Harvey:2014qla,MeurenLCFA,Ilderton:2018nws}, as well as recent comparisons of theory with laser experiments~\cite{Cole:2017zca,Poder:2018ifi}, have highlighted shortcomings of existing models and simulations. A  priority of the research field as a whole is hence the development of the right approximations and tools to more accurately model strong-field processes analytically and numerically. This requires a fresh look at previously neglected processes.

The analytical study of strong-field QED processes {in intense laser} fields is mostly devoted to those in which {a single} incoming particle, typically a high-energy electron or photon, {scatters off the field}, producing other photons and/or electron-positron pairs. A natural direction in which to extend existing efforts is then to consider processes involving multiple incoming particles. Two such processes are pair-annihilation to one photon, and one-photon absorption by an electron. {(See e.g.~\cite{zhukovskii73,dipiazza08b,Hu:2011eq,gies14,king15c,Vorosh15,gies16,Hartin:2017uah} for other processes.})  To our knowledge {absorption} has not been considered in detail for \textit{pulsed} laser fields (though for the constant field case see~\cite{RitusReview}, and below). Although the process appears simple at first sight, we will show that one can learn a lot about the regularisation of strong-field processes and the applicability of the locally constant field approximation (LCFA)~\cite{RitusReview} which lies behind ``particle in cell'' (PIC) simulations of QED effects in laser-plasma interactions, for a review see~\cite{Gonoskov:2014mda}.

This paper is organised as follows. In Sect.~\ref{sec:notation} we state our conventions and briefly recap the basics of scattering calculations in strong field QED. In Sect.~\ref{sec:cross} we provide some general results on the probability and cross section of scattering processes in strong fields, which we then apply to the particular process of photon absorption by an electron in a laser pulse, in Sect.~\ref{sec:absorption}. Here we discuss the subtle regularisation of the process, and the relation to gauge invariance, as well as the physical interpretation of the process, and its perturbative, soft, and collinear limits. With an eye to future numerical simulation of photon absorption, we consider {its LCFA} in Sect.~\ref{sec:LCFA}. We find notable differences in the behaviour of the LCFA when compared with other {processes. The LCFA} is benchmarked against exact calculations, and we discuss its regime of validity. We conclude in Sect.~\ref{sec:concs}.

\subsection{QED in plane wave background fields  \label{sec:notation}}

We consider QED in the presence of a background plane wave, described by the (scaled) potential $e A_\mu(x) = a_\mu(\varphi)$ depending only on the single dimensionless variable $\varphi = k . x$ where $k_\mu$ is a lightlike wave vector. We use lightfront coordinates such that an arbitrary vector $x^\mu$ has components $x^\pm = x^0 \pm x^3$ and $\bm{x}^\perp \in \{x^1,x^2\}$, the latter referred to as the ``transverse'' coordinates. Covariant vectors are $p_\LCpm = (1/2)(p_0 \pm p_3)$ and $\bm{p}_\LCperp \in \{p_1,p_2\}$. We can always choose coordinates such that $k_\mu = k_\LCp \delta_\mu^\LCp$. The potential $a_\mu$ can then be chosen to have only transverse components and obey $a_\mu(-\infty)=0$, such that  its derivative gives the electric field~\cite{Dinu:2012tj}. We consider pulsed fields, i.e.~those having finite duration in $\varphi$ or being asymptotically switched off for large $|\varphi|$. Classically, an electron of initial momentum $p_\mu$ entering the wave has the subsequent kinematic momentum
\be\label{kinetic}
\pi_p^\mu(\varphi) = p^\mu -a^\mu(\varphi) + k^\mu \frac{2p . a(\varphi)-a(\varphi)^2}{2k. p}	\;,
\ee
which is the exact solution of the Lorentz force equation in the background. Turning to QED, the effect of the background on charged particles is treated exactly through the use of the Volkov wavefunctions~\cite{volkov35}. For an electron of initial momentum $p_\mu$ interacting with the plane wave we define the Volkov wavefunction  $\Psi_p$ by
\begin{align}
\Psi_p
=&
\left(
1
+
\frac{\slashed{k} \slashed{a}(\varphi)}{2 k . p}
\right)
u_p
e^{
	- 
	i p . x
	-
	i S_p(\varphi)
}
\;, \qquad
S_p(\varphi) :=
\int\limits^{\varphi} \!\ud \phi\, 
\frac{2 p . a(\phi) - a^2(\phi)}{2 k . p}
\label{volkov}
\end{align}
Working in the Furry picture~\cite{Furry51}, these functions represent the external legs (asymptotic particles) in scattering amplitudes. For more details see the recent review~\cite{Seipt:2017ckc}.


\section{Cross sections in background plane waves \label{sec:cross}} 
For strong field QED processes with a single particle in the initial state, a natural observable is the total probability of a process, as we review below. However, for processes with two incoming particles, where the interaction is fundamentally collisional, it is more standard to consider cross-sections. Here we extend some textbook results for {QED without background} to the analogous situation when a {background plane wave is} present.

Any $S$-matrix element $S_{fi}$ in a plane wave background, calculated using the asymptotic wavefunction (\ref{volkov}) (and the corresponding propagator) has the structure
\be
S_{fi} = (2\pi)^3 \delta^3_\lf(p_f - p_i) M_{fi} \;,
\label{smatrixLF}
\ee
in which the delta function $\delta^3_\lf(p) \equiv \delta_\LCm(p)\delta^2_\LCperp(p)$ conserves overall momentum in \text{three} directions, and where $M_{fi}$ (the invariant matrix element as defined in e.g.~\cite{Peskin:1995ev}) has mass dimension $3-N_{i}-N_{f}$ for $N_{i}$ and $N_{f}$ the number of initial and final state particles respetively. Taking the mod-square we have 
\be\label{TillP1}
|S_{fi}|^2  =V_\lf (2\pi)^3 \delta^3_\lf(p_f-p_i) |M_{fi}|^2 \;,
\ee
in which $V_\lf \equiv (2\pi)^3\delta^3_\lf(0)$ is the lightfront volume. The appearance of this factor is particular to plane wave backgrounds. {We then include the state normalisation factor  
	\be\label{TillP2}
	\prod_i \frac{1}{2E_i V} \prod_f \frac{1}{2 E_f V} \;,
	\ee
	and to obtain the final probability we sum/average over spins and integrate over outgoing momenta with the standard measure}
\be\label{TillP3}
\prod_f \frac{V}{(2\pi)^3}\ud^3{\bm{p}}_f \;.
\ee
Note that $V$ is the standard Cartesian volume, because the initial and final states are defined at $x^0\to \pm \infty$, and must be independent of the specific background field configuration we consider. We will resolve the presence of two volume factors below. Combining (\ref{TillP1})--(\ref{TillP3}) gives the total scattering probability as
\be\label{Pallm}
\mathbb{P} =  \int\prod_f \frac{\ud^3{\bm{p}}_f}{(2\pi)^3 2E_f} \frac{V_\lf (2\pi)^3 \delta^3_\lf(p_f-p_i) |M_{fi}|^2}{\prod\limits_i {2 E_i V}} \;.
\ee
One may verify by power counting that this is dimensionless, as it should be. We now consider particular choices for the number of incoming particles.

\subsection{One incoming particle }
Consider any process initiated by a single particle, momentum $p_\mu$, with any number of final state particles. Applying the general result (\ref{Pallm}) it remains to understand the ratio of volume factors $V_\lf/V$. Now, we note that the quantity $2E_p V$ is just the initial state normalisation, and is by construction Lorentz invariant~\cite{Peskin:1995ev}, being $2E_p \delta^3({\bm{p} - \bm{q}})$ in the limit $\bm{q} \to \bm{p}$. It can easily be checked by changing variables in this expression that
\be
2E_p \delta^3({\bm{p} - \bm{q}}) = 2p_\LCm \delta_\lf(p- q) \implies 2E_p V = 2p_\LCm V_\lf \implies \frac{V_\text{LF}}{2p_0 V} = \frac{1}{2p_\LCm} \;.
\ee
This allows us to attribute a physical meaning to the ratio of volume factors, which cancel, yielding the final expression for the total scattering probability as
\be
\mathbb{P} = \frac{1}{2 p_\LCm} \int\prod_f \frac{\ud^3{\bm{p}}_f}{(2\pi)^3 2E_f} (2\pi)^3 \delta^3_\lf(p_f-p) |M_{fi}|^2 \;.
\ee
It is important to stress that this is \textit{precisely} the same result as is obtained by starting with properly normalised wavepackets~\cite{Ilderton:2012qe}, and, at the end of the calculation, taking those wavepackets to have the usual narrow spread. In particular this justifies our manipulations of the infinite volume factors, which yields the correct leading factor of $1/2p_\LCm$ in the probability~\cite{Ilderton:2012qe}. 

\subsection{Two incoming particles: the cross section }
Now consider the case of two initial state particles, momenta $p_\mu$ and $l_\mu$, and again any number of final state particles. In this case it is more common to consider the cross section. This is as usual defined by
\be\label{sig-def}
\sigma = \frac{\text{transition probability per unit time}}{\text{flux}}  = \frac1{\text{flux}} \frac{\mathbb{P}}{T} \;,
\ee
with $\mathbb P$ from (\ref{Pallm}) and the two-particle flux factor is, assuming for example that $l^2=0$ and $p^2=m^2$ as we will consider below~\cite{Peskin:1995ev},
\be\label{flux}
\frac{1}{\text{flux}} = \frac{V E_l E_p}{l.p} \;.
\ee
Combining (\ref{sig-def}) and (\ref{flux}) we have
\be
\sigma = \frac{1}{4l.p}  \frac{V_\lf}{VT} \int\prod_f \frac{\ud^3{\bm{p}}_f}{(2\pi)^3 2E_f} \, \, (2\pi)^3 \delta^3_\lf(p_f - p - l )  \,\, |M_{fi}|^2 \;.
\ee
Again we must deal with the volume factors. To deal with two incoming particles, it was suggested in \cite{RitusReview} to use methods similar to those for a single incoming particle. {This requires treating the incoming particles asymmetrically, which may} be appropriate for specific scenarios~\cite{Us3}, but here we give a more general method. The volume factors may be manipulated using the relation $VT = (1/2)V_\lf T_\lf$, {which serves to define the lightfront time $T_{\lf}$}, where the $1/2$ is the Jacobian determinant for changing variables to lightfront coordinates. With this we obtain
\be
\sigma = \frac{1}{2l.p T_\lf}  \int\prod_f \frac{\ud^3{\bm{p}}_f}{(2\pi)^3 2E_f} \, \, (2\pi)^3 \delta^3_\lf(p_f - p - l )  \,\,  |M_{fi}|^2 \;.
\label{cross2}
\ee
The presence of $T_\lf$, the interaction duration in $x^\LCp$ rather than in $x^0$, is natural due to the form of the background. This will be, in the example below, simply the spacetime extent, {in $x^\LCp$, of} the background field.


\section{Properties of the absorption cross section \label{sec:absorption}}
%
Having established the general expression (\ref{cross2}) for the cross section in a plane wave background,  we turn to the explicit example of photon absorption by an electron. Let the electron have initial momentum $p_\mu~(p^2 = m^2)$, and the photon have momentum, $l_\mu~(l^2 = 0)$. The final state electron momentum is $p^\prime_\mu$. This process is kinematically forbidden in vacuum, but allowed in a background.

The $S$-matrix element $S_{fi}$ for absorption is, in terms of the Volkov wavefunctions (\ref{volkov}), 
\begin{align}
S_{fi}
=&
-
i
e
\int\! \ud^4 x\, 
\bar{\Psi}_{p^\prime}
\slashed{\varepsilon}
\Psi_p
e^{-i l . x}
\;,
\label{smatrix}
\end{align}
where $\varepsilon_\mu$ is the polarisation vector of the incoming photon and $\slashed{\varepsilon} \equiv \gamma^\mu \varepsilon_\mu$  for  $\gamma^\mu$ the Dirac matrices. The $S$-matrix element is readily expressed in the form (\ref{smatrixLF}) with $p_i = p + l$, $p_f = p ^\prime$ and
\begin{align}\label{mfi2}
M_{fi}
=&
-
\frac{1}{2}
ie
\int\! \frac{\ud\varphi}{k_+}\, 
e^{
	i \Phi(\varphi)
}
\text{Spin}(a)
\;,
\end{align}
where we have defined the spin structure
\begin{align}\label{spin}
\text{Spin}(a)
= 
\bar{u}_{p^\prime}
\left(
1
+
\frac{\slashed{a}(\varphi) \slashed{k}}{2 k . p^\prime}
\right)
\slashed{\varepsilon}
\left(
1
+
\frac{\slashed{k} \slashed{a}(\varphi)}{2 k . p}
\right)
u_p \;,
\end{align}
and phase term,
\begin{align}\label{phaseterm}
{\Phi(\varphi)
	=
	-\int\limits^\varphi  \!\ud \phi\,
	\frac{l. \pi_p(\phi) }{k.(p+l)}
	\;.}
\end{align}
The $S$-matrix element (\ref{smatrix}) {becomes identical to that of nonlinear Compton scattering if one identifies $-l_\mu$ ($\varepsilon_\mu^*$) with the outgoing photon momentum (polarisation) in that process}. The key differences in the two processes enter at the level of the probability (or cross section), due to the different number of incoming and outgoing particles. Here, there will be no freedom in the final electron momentum, as it will be entirely determined by the incoming momenta through the delta-function in (\ref{smatrixLF}). The fact that we only have one outgoing particle also has an impact on the  regularisation of the $S$-matrix element, which is the next step of the calculation.

\subsection{Regularisation and gauge invariance } \label{sect:reg}
An ambiguity now arises in the calculation of the probability; different results are obtained depending on whether one uses Lorentz gauge or lightfront gauge. The difference is a boundary term which can be traced back {(as is well-known for e.g.~the crossing-symmetric processes of absorption such as nonlinear Compton scattering), to a term in the pre-exponential integrand of (\ref{mfi2})} which is independent of $a_\mu$. This term constitutes an integral over a pure phase, the definition of which is ambiguous. {If the laser pulse has support only on the finite interval $\varphi\in\{\varphi_i, \varphi_f\}$ then it is the apparent contributions from outside the pulse, where there is no field, which cause problems.} One way to make the integral better defined, and manifestly convergent, was found in~\cite{Boca:2009zz} in the context of nonlinear Compton scattering. One introduces convergence factors $\exp(-\epsilon |\varphi|)$ to the integrand of $M_{fi}$ as follows,
\be\label{SFI-REG}
\int\!\ud\varphi\, e^{i\Phi(\varphi)} \, \text{Spin}(a) 
\longrightarrow 
\int\limits_{-\infty}^{+\infty}\!\ud\varphi\, e^{i\Phi(\varphi)} \, \text{Spin}(a) \, e^{- \epsilon |\varphi|}
,
\ee
{which cuts the integration into three parts corresponding to before, during and after the pulse. Integrating by parts, one finds that contributions from outside the pulse are excised, while the} phase integral becomes restricted to the pulse duration, with a modified integrand; explicitly
\be\label{SFI-REG2}
\int\limits_{-\infty}^{+\infty}\!\ud\varphi\, e^{i\Phi(\varphi)} \, \text{Spin}(a) \, e^{- \epsilon |\varphi|}
\to
-
\int\limits_{\varphi_i}^{\varphi_f}\!\ud\varphi\, e^{i\Phi} \frac{\ud}{\ud\varphi}\bigg(\frac{\text{Spin}(a)}{i\Phi'}\bigg)
,
\ee
where a prime denotes differentiation with respect to the laser phase. Note that $\Phi'$ cannot vanish. The expression (\ref{SFI-REG2}) also holds for the case of pulses with DC components, which exhibit a memory effect~\cite{Dinu:2012tj}, or for asymptotically switched fields, in which case $\varphi_i=-\infty$ and $\varphi_f=+\infty$. The reason that this is a correct method to use (aside from the fact that (\ref{SFI-REG2}) is unambiguous, finite, and has the correct zero-field limit) is that it makes the gauge invariance of the scattering amplitude \textit{explicit}~\cite{Ilderton:2010wr,Dinu:2012tj}. As a result, when calculating the probability, calculations in lightfront and Lorentz gauge agree exactly.  For applications of this method to various processes see~\cite{Ilderton:2010wr,Seipt:2010ya,Mackenroth:2010jr,Mackenroth:2018smh}, and for comparisons of calculations in Lorentz/lightfront gauges see~\cite{Dinu:2017uoj,Dinu:2018efz}.

The presence of the phase derivative $\Phi'$ (see (\ref{phaseterm})) in the  denominator of (\ref{SFI-REG2}) complicates final state integrations, however, and it would be useful to have a simpler, equivalent expression before proceeding. For pulses with no DC components the regulated expression in (\ref{SFI-REG2}) is exactly equal to
\be\label{reg}
-
\int\limits_{\varphi_i}^{\varphi_f}\!\ud\varphi\, e^{i\Phi} \frac{\ud}{\ud\varphi}\bigg(\frac{\text{Spin}(a)}{i\Phi'}\bigg)
=
\int \!\ud\varphi \, e^{i\Phi} \bigg[ \text{Spin}(a) -\frac{l.\pi_p(\varphi)}{l.p}\text{Spin}(0) \bigg]\;.
\ee
where $\text{Spin}(0)$ denotes the part of the spin term (\ref{spin}) which is independent of $a_\mu$. The equivalence of (\ref{reg}) and (\ref{SFI-REG2}) can be shown explicitly by separating $\text{Spin}(a)$ into field-dependent and independent pieces, integrating by parts, and using the definition of the kinetic momentum (\ref{kinetic}).  {The preceding issues are well-understood for other processes, but are sometimes neglected and implicitly resolved by another method; as we now discuss, the absorption process does not allow this.}

\subsection{Cross section }
The remainder of the calculation of the cross section (\ref{cross2}) is straightforward and follows standard {methods. The} final-state electron momenta can be integrated out trivially using overall momentum conservation. This means that the final electron momentum $p'_\mu$ is specified exactly by the initial momenta. Explicitly, this is
\be
p'_\mu = p_\mu + l_\mu - \frac{l . p}{k . (l+p)} k_\mu \;.
\ee
For the regulating factors in (\ref{reg}) we define
\be\label{reg-rep}
\Delta := 1-\frac{l.\pi_p(\varphi\big)}{l. p} \;,
\ee
and we write $\Delta^{\!\prime}$ for the same factor depending on $\varphi^{\prime}$, the phase argument of the conjugate $S$-matrix element. Then the final expression for the absorption cross section is
        
\begin{align}\label{SIGMA0}
\nonumber \sigma = \pi \alpha \frac{m^2}{l.p} \frac{1}{T_\text{LF}}\,\frac{1}{k . p (1 + s)}\int\! \ud\varphi\!\int\!\ud \varphi'
& \cos\left[ (\varphi-\varphi')
\frac{l. \langle\pi_p\rangle}{k . p (1 + s)}
\right] \\
&\Big(	\frac{g(s)}{m^2} \Big[a^2(\varphi)\Delta^{\!\prime} + a^2(\varphi')\Delta - 2 a(\varphi) . a(\varphi') \Big] - \Delta \Delta^{\!\prime}
\Big) \;,
\end{align}
{in which, and from here on, we absorb a factor of $k_\LCp$ into $T_\text{LF}$ which becomes the dimensionless phase length of the field} and we define a function $g(s)$ and an average $\langle \,.\, \rangle$ by   
\be
g(s) := 
\frac12
+
\frac{s^2}{4(1 + s)}
\;,
\qquad
\langle f \rangle := \frac{1}{\varphi-\varphi'} \int\limits_{\varphi'}^{\varphi} \!\ud \tilde{\varphi}\,  f(\tilde{\varphi}) \;,
\label{g}
\ee
where $s = k.l/k.p$ is the lightfront momentum fraction of the incoming particles. Due to the regularising factors, every term in the integrand of (\ref{SIGMA0}) vanishes outside the pulse envelope, which reflects the physical result that the absorption process does not occur without the presence of the background. Notice the asymmetry in the $a^2$ terms; for example $a^2(\varphi)$ clearly gives a convergent $\varphi$ integral only, hence a factor $\Delta^{\!\prime}$ appears to force the $\varphi'$ integral to converge. No such factors appear in the $a(\varphi) . a(\varphi')$ term, but here none are needed, so this is consistent.

It is interesting to compare this with the expression obtained \textit{without} explicitly regularising the amplitude. This is conveniently expressed in terms of the sum and difference of phase variables, $\phi := \frac{1}{2} (\varphi + \varphi')$ and $\theta :=  \varphi - \varphi'$ as 
\begin{align}\label{SIGMA}
\sigma = \pi \alpha \frac{m^2}{l . p} \frac{1}{T_\text{LF}} \frac{1}{k. p(1 + s)}
\int\! \ud\phi\! \int\!\ud \theta
\cos\left[ \theta
\frac{l.\langle\pi_p\rangle}{k. p(1+s)}
\right]
\Big(
\frac{g(s)}{m^2}
\theta^2
\langle
a^\prime
\rangle^2
-1
\Big)  + \# \text{total derivative}\;.
\end{align}
In this expression the final term is a total derivative, the coefficient of which is dependent on the chosen gauge. In calculations of, say, nonlinear Compton where there are two or more outgoing particles, such boundary terms are dropped, and an alternative regularisation is used; performing the Gaussian integrals over the final state \textit{transverse} momenta~\cite{Dinu:2013hsd} introduces $1/(\theta + i \epsilon)$ factors, the imaginary part of which has the same effect as the procedure in (\ref{sect:reg}) and the factors (\ref{reg-rep}), namely to subtract the `zero field' contributions and make {the integrals manifestly convergent}~\cite{Dinu:2013hsd}. Here, however, because the momentum of the single outgoing particle is fixed, this method is not {available, and so the regularisation factors $\Delta$ in (\ref{SIGMA0}) must be retained.}

\subsection{Soft and collinear limits } \label{SECTION:SOFT}
As our process involves the absorption of massless particles we are prompted to consider, {here and in the section below}, possible infra-red divergences and degenerate processes. We consider first the soft limit in which the incoming photon energy goes to zero, $l_0\to 0$. Taking this limit is made easy by our identification of the regularising structures~(\ref{reg-rep})--(\ref{SIGMA0}). Under the integrals of either the probability or cross section (\ref{SIGMA0}), we can set $l_0=0$ directly (implying $s=0$) without introducing any divergence. The  $(\varphi,\varphi')$ integrals are then Fourier transforms, at zero frequency, of either $a_\mu$ or $a^2$. We have assumed that the wave is not unipolar, so the former integral is zero. This implies that the Fourier transform of $a^2$ at zero frequency is then finite, and it follows that the integral in the cross section gives a finite result. Now observe that the prefactor, again in either the probability or the cross section, diverges like $1/l_0$ as $l_0\to 0$. This is hence the leading order soft behaviour. Note also that integration over an initial photon momentum distribution $f(l)$ with support far from $l_0=0$ would then produce an infra-red finite result, whereas if the wavepacket had `flat' support $f(l_0)\sim  \text{constant}$ for $l_0 \sim 0$, then we would reproduce the usual log divergence of QED~\cite{Yennie:1961ad,Peskin:1995ev}.

We consider also a collinear limit, in which the incoming photon is parallel to the laser. In nonlinear Compton this limit is finite, but it is subtle to obtain~\cite{MeurenLCFA} {when using the ``$\theta + i \epsilon$'' regulated expression}. Taking the same limit for photon absorption is much simpler {because of the \emph{explicit} presence of $\Delta$ in}~(\ref{SIGMA0}); as $l_\mu \to (\omega_l/\omega) k_\mu$ for some frequency $\omega_l$, the regularising factor $\Delta$ vanishes, since then $l.\pi \to l.p$. Hence the only term we need consider is the cross-term in (\ref{SIGMA0}) with no $\Delta$ factors. In this term $g(s) \to g(0)=1/2$, the exponential terms reduce to Fourier transform kernels, and the phase integrals can then be performed immediately to yield 
{
	\be
	\label{sigma-limit}
	\sigma \to \frac{\alpha \pi}{l.p ~k.p~ T_\text{LF}}\, |\tilde{a}_\LCperp(\omega_l/\omega)|^2 \;,
	\ee
}
in which $\tilde{a}$ is the Fourier transform of the potential {with respect to $\varphi$}. Hence the collinear absorption cross section (and probability) take a particularly simple form, similar to the corresponding final result for nonlinear Compton~\cite{MeurenLCFA}.

\subsection{Physical interpretation }

Some insight into the absorption process is given by examining its weak-field limit, $a_0\ll 1$. It is simplest to do so at the level of the amplitude; we take (\ref{smatrix}) and expand directly in powers of the background field, retaining only the lowest order contribution, which is linear in $a_\mu$. Ignoring irrelevant prefactors, and using some simple manipulations involving standard $u$-spinor identities, the result may be written as 
\be\label{pert-compton}
S_{fi} \sim \int\!\ud \omega' \delta^4(p'+k'-p-l) \bigg[{\bar u}_{p'}\slashed{a}(k') \frac{\slashed{p}+\slashed{l}+m}{2l.p}\slashed{\varepsilon}u_p - {\bar u}_{p'}\slashed{\varepsilon}\frac{\slashed{p}-\slashed{k'}+m}{2k'.p}\slashed{a}(k') u_p \bigg] \;,
\ee
in which $a_\mu(k')$ is the Fourier transform of the background so $k'_\mu= (\omega' /\omega) k_\mu$, c.f.~(\ref{sigma-limit}). The structure of this perturbative result is clear; it is the textbook expression for tree level Compton scattering, convoluted with the intensity profile of the external field. This is as expected. What is unusual is that the intensity profile is attached to the \textit{outgoing} photon (hence, as one may check using the delta function, only the negative, or outgoing, frequency modes of the field contribute).  This is in contrast to the perturbative limits of strong field processes with a single incoming particle, such as nonlinear Compton scattering, trident, and nonlinear Breit-Wheeler, where the background acts as a source of \textit{initial state} photons, effectively stimulating a process which is kinematically forbidden in vacuum.

\begin{figure}[t!]
	\includegraphics[width=0.75\textwidth]{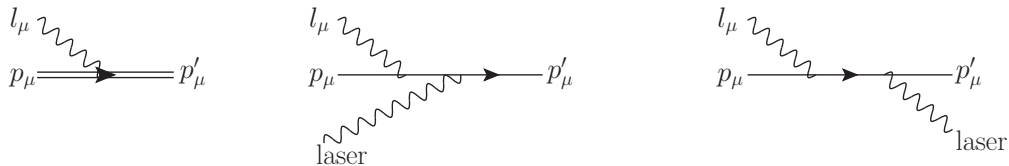}
	\caption{\label{FIG:PERT} \textit{Left:} {The Furry picture diagram for absorption, in which the double lines here represent the Volkov wavefunctions. \textit{Middle}:} Lowest-order ``laser-stimulated'' absorption, in which laser photons in the initial state are absorbed; this diagram vanishes by momentum conservation. \textit{Right}: the actual perturbative limit of the absorption process, in which the negative frequency part of the field is sampled, i.e~the perturbative limit is Compton scattering in which the outgoing photon is convoluted with the laser profile.}
\end{figure}

The absorption amplitude behaves differently. Naively, the physical picture one might have in mind corresponds to the middle diagram in Fig.~\ref{FIG:PERT}; the incoming ``probe'' photon and electron are accompanied by a laser photon, the presence of which allows the probe photon to be absorbed. However, this process and all similar processes with higher numbers of initial state laser photons are also kinematically forbidden. The lowest order term (\ref{pert-compton}) instead corresponds to the right hand diagram in Fig.~\ref{FIG:PERT}, which is Compton scattering. In this way it looks as if the laser background is `generated' by the scattering process, which is clearly not the correct physical picture.  

Observe that such contributions are implicitly included in all amplitudes calculated in background plane waves; because these backgrounds are ``on-shell'', they can be transformed into asymptotic coherent states which are present in both the in-state and out-state~\cite{Kibble:1965zza,Frantz,Gavrilov:1990qa}, hence the perturbative expansion of any process yields diagrams like that on the right of Fig.~\ref{FIG:PERT}, it is just that for e.g.~nonlinear Compton, such terms appear at higher orders in the perturbative expansion. What distinguishes photon absorption from other processes is that, in both the perturbative limit and the exact $S$-matrix element, the \emph{net} amount of laser energy (or rather lightfront energy) playing a role in the interaction must be negative. In other words, for the process to occur there must actually be a net \textit{emission} of photons. 

{This prompts an} alternative interpretation of the lowest order contribution to absorption. When asking for the total probability of absorption without emission in a plane wave background, {which is what the corresponding Furry picture diagram in Fig.~\ref{FIG:PERT} represents, one should sum over degenerate (indistinguishable) final states in order to have a better-defined, inclusive observable. This sum} this would naturally include unobservable emission \textit{into} that part of momentum space corresponding to the background spectrum; this is just the kind of contribution which is being picked up in the perturbative limit~(\ref{pert-compton}).


\section{Locally constant field approximation \label{sec:LCFA}}

The zero frequency limit of a monochromatic plane wave gives a constant ``crossed'' field. Scattering processes in such fields provide a bridge between plane wave calculations and experiment through numerical simulation via particle-in-cell codes, see~\cite{Gonoskov:2014mda} for a review. Such codes use the \text{locally} constant field approximation (LCFA) to adapt the constant crossed field results to arbitrary field structures, on the basis that over the relevant QED timescales any background field is \emph{instantaneously} constant~\cite{RitusReview}. Such an approach is powerful, but has shortcomings. For single-vertex processes, the LCFA fails at low energy~\cite{Dinu:2012tj}, at low lightfront momentum~\cite{Harvey:2014qla,MeurenLCFA,Ilderton:2018nws}, at very high energy~\cite{Podszus:2018hnz,Ilderton:2019kqp} and in the presence of strong field gradients \cite{King:2019cpj}. With an eye to the future numerical implementation of absorption, we now turn to the properties and limitations of its LCFA.

The LCFA of a given process may be calculated by specialising to the constant crossed field case and then ``localising'' variables depending on the field amplitude. An alternative and more revealing approach is to start with a result valid in a general plane wave, and then expand in a suitable parameter such that the LCFA is obtained~\cite{MeurenLCFA,Ilderton:2018nws}.  Essentially, one makes a ``short coherence time'' approximation of e.g.~the cross section (\ref{SIGMA}) by expanding in powers of the phase variable $\theta \ll 1$, which is related to the coherence time of the process.  For e.g.~nonlinear Compton scattering, this expansion is most commonly and easily performed at the level of the integrated probability. Here, though, the lack of final state integrals means that what we are essentially trying to construct is an angularly resolved LCFA, see~\cite{DiPiazza:2018bfu, Blackburn:2019lgk} and references therein. We begin with (\ref{SIGMA0}). In order to match the constant crossed field result it is necessary to treat different terms in different ways, as dictated by their dependence on $\theta$~\cite{TackaSeipt}. One first writes, in (\ref{SIGMA0}),
\begin{align}
l.\langle \pi_p \rangle
=&
\frac{s}{2}
m^2
\mu
-
\frac{1}{2s}
(l - s \langle\pi_p\rangle)^2
,
\label{exponent}
\end{align} 
in which Kibble's effective mass is~\cite{Kibble:1975vz,Harvey:2012ie},
\begin{align}
\mu \equiv
\frac{\langle \pi_p\rangle^2}{m^2}
= 
1
+
\frac{\langle {\bm{a}}_\perp^2\rangle}{m^2}
-
\frac{\langle {\bm{a}}_\perp\rangle^2}{m^2}
.
\label{kibble}
\end{align}
The Kibble mass (\ref{kibble}) contains the second moment of the kinetic momentum (and consequently the laser vector potential), and as such its leading order non-trivial short coherence time approximation comes in at $\mathcal{O}(\theta^2)$,
\begin{align}
\mu
\simeq
1
-
\theta^2
\frac{(a^\prime(\phi))^2}{12 m^2} \;.
\label{kibbleapprox}
\end{align}
All other terms appearing in the cross section instead have leading non-trivial approximations at $\mathcal{O}(1)$, 
\be
\langle\pi_p\rangle
\simeq 
\pi_p(\phi) \;,
\qquad
\langle a^\prime\rangle
\simeq
a^\prime(\phi)
\;, 
\label{mean}
\ee
which are simply the mean values.  Applying these approximations, the $\ud\theta$ integral in (\ref{SIGMA0}) is reduced to Airy form, and the LCFA is obtained.  Precisely the same result is obtained by starting with the constant crossed field results (for which $a_\mu(\phi) \rightarrow \tilde{a}_\mu \phi$ with $\tilde{a}_\mu$ a constant polarisation vector), justifying the approximations (\ref{kibbleapprox}) and (\ref{mean}).  Explicitly, the LCFA result is
\begin{align}
\sigma_{LCFA}
=
\frac{4 \pi^2 \alpha}{l . p}
\frac{1}{T_{LF}}
\int 
\ud\phi \,
\frac{1}{s}
(
4
g
\bar{z}
-
z
)
\Ai(\bar{z}) \;,
\label{crossLCFA}
\end{align}
in which the function $g \equiv g(s)$ is as in (\ref{g}) and the argument of the Airy function is
\begin{align}
\bar{z}(\phi)
=
2z(\phi)
\frac{l . \pi_p(\phi)}{s m^2} \;,
\qquad 
z(\phi)
=
\bigg(
\frac{1}{\chi_e(\phi)}
\frac{s}{1+s}
\bigg)^{2/3} \;,
\label{zbarz}
\end{align}
where the (local) quantum nonlinearity parameter for the electron is 
\begin{align}
\chi_e(\phi) \equiv \frac{\sqrt{p^\mu F^2_{\mu\nu}(\phi) p^\nu}}{m^3} \;,
\end{align}
and the corresponding parameter for the photon reduces to $\chi_\gamma(\phi) = s \chi_e(\phi)$.

Comparison with the well known LCFA (i.e. constant crossed field) expressions of nonlinear Compton scattering \cite{nikishov64,kibble64} and nonlinear Breit-Wheeler~\cite{nikishov67,narozhny69} shows {a significant difference}. The LCFA for those processes has a nontrivial dependence only on the $\chi$-parameters of the involved particles. Here, though, the integrand of (\ref{crossLCFA}) depends not only on $\chi_e$ but also on the local kinetic momentum ${\pi^\mu_p}$ of the electron, which appears in $\bar{z}$.

That individual dependencies on intensity and energy, as seen in general pulsed expressions, are traded for a dependence only on $\chi$ in the LCFA, is partly due to the nature of the approximation but also partly due to being able to perform the Gaussian integrals over outgoing perpendicular momenta. Since these integrals are absent for absorption the situation is different, and there persists a dependence on local parameters other than just $\chi$.  LCFA expressions are thus not, in general, dependent only on $\chi$.  This has consequences which we discuss in Sect.~\ref{SECTION:WINGS}.

We comment that, as for the general results above, there is no photon-laser collinear divergence in the LCFA. The soft behaviour of the LCFA is different; aside from the $1/l_0$ dependence in the prefactor of (\ref{crossLCFA}) the \textit{integrand} behaves as $\sim 1/l_0^{1/3}$ as $l_0\to 0$; this is typical of LCFA results, see~\cite{Dinu:2012tj,MeurenLCFA,Ilderton:2018nws} for discussions.

\subsection{Comparison: LCFA vs monochromatic}
%
%
\begin{figure}[t!]
	\includegraphics[width=0.47\textwidth]{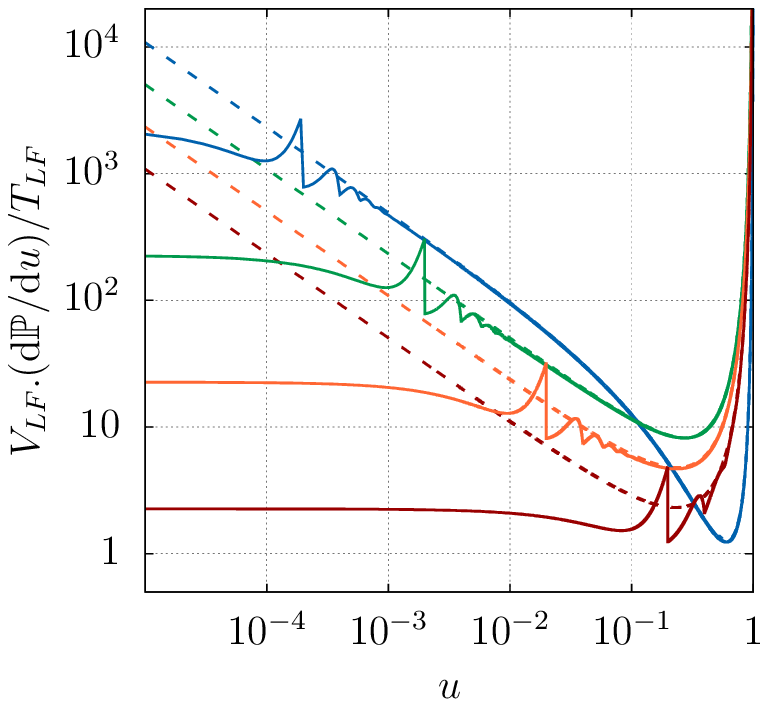}
	\includegraphics[width=0.47\textwidth]{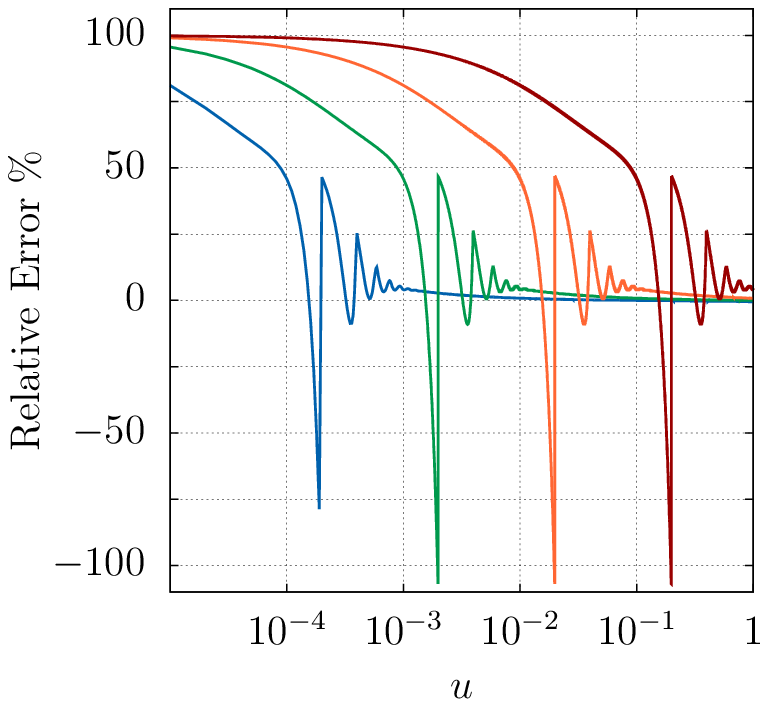}
	\caption{\label{FIG:MONO1} \textit{Left:} Comparison of the differential emission probability in a monochromatic wave following the prescription (\ref{exactmono}) (solid lines) and the locally constant field approximation (LCFA, dashed lines) following the prescription (\ref{lcfamono}):  $a_0=10$ is fixed, and $k.p/m^2 = 0.01$ (blue), $0.1$ (green), $1$ (orange), $10$ (red). As with other strong field QED processes, the LCFA misses the harmonic structure of the monochromatic case at small $u$.  While the approximation is better at larger $u$ at fixed parameters, the LCFA is overall worse as energy increases. 
		\textit{Right:} Relative error between LCFA and exact monochromatic results.}
\end{figure}
\begin{figure}[t!]
	\includegraphics[width=0.47\textwidth]{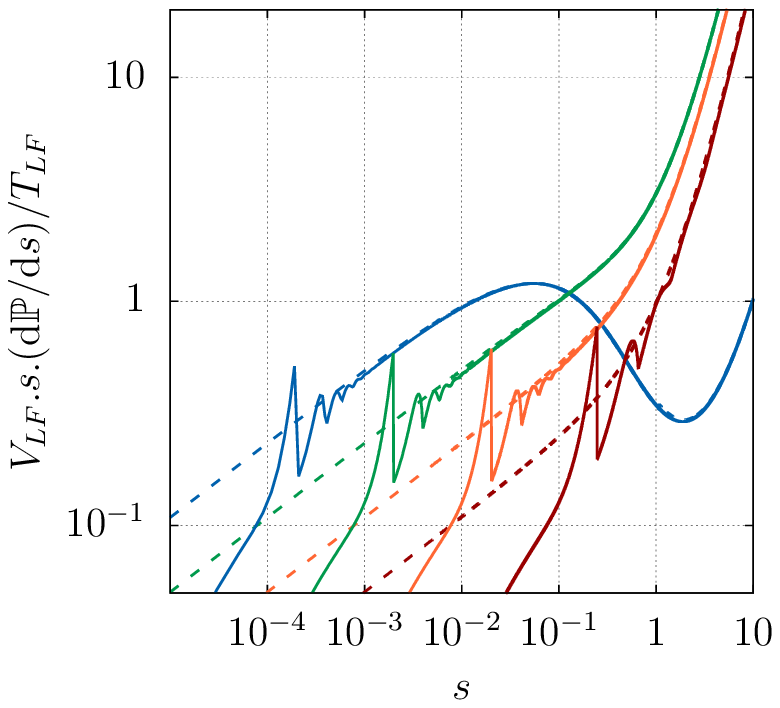}
	\includegraphics[width=0.47\textwidth]{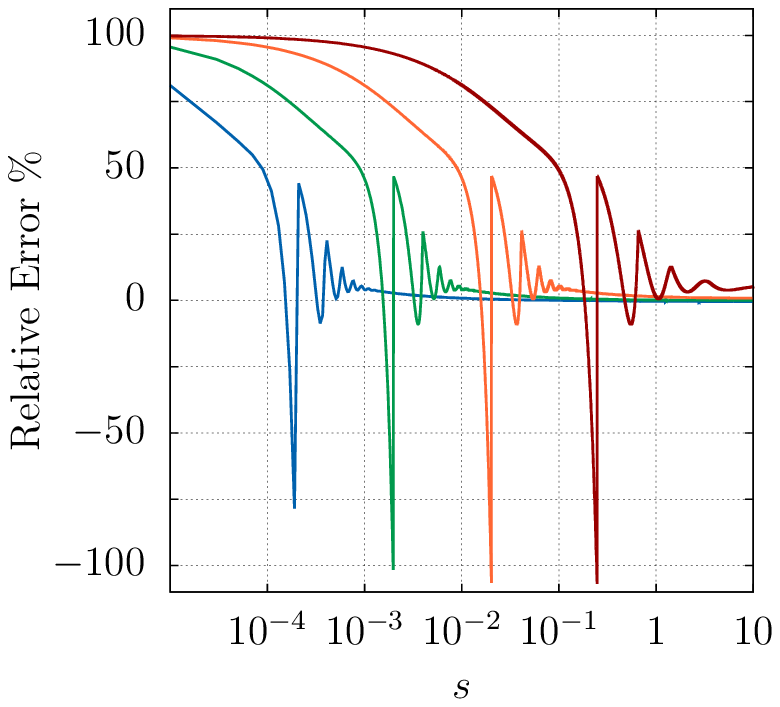}
	\caption{\label{FIG:MONO2}  \textit{Left:} Comparison of the differential ``energy'' spectrum $s \ud \mathbb{P}/\ud s$ in a monochromatic wave following the prescription (\ref{exactmono}) (solid lines) and the locally constant field approximation (LCFA, dashed lines) following the prescription (\ref{lcfamono}):  $a_0=10$ is fixed, and ${b_0 = }k.p/m^2 = 0.01$ (blue), $0.1$ (green), $1$ (orange), $10$ (red). As above, the LCFA is more accurate for larger values of $s$, and generally for lower $b_0$ at fixed $a_0$. For $b=0.01$ curve, the electron's quantum nonlinearity parameter $\chi_e$ is less than one, and there is marked change in the behaviour of the spectrum for higher $s$ {(see main text for details)}. \textit{Right:} Relative error between LCFA and exact monochromatic result. {For all cases except $b_0 = 10$ the LCFA converges with the exact result for $s \gtrsim b_0$. For $b_0 = 10$, further increasing $s$ the error begins to converge to a small, but non-zero error. This is due to a failure of the LCFA for larger values of $b_0$.}} 
\end{figure}
In order to check the quality of the LCFA we need analytic results against which to compare. A common approach is to benchmark against results in a monochromatic wave, for which well-understood exact results are available. However, for absorption, because of the particular nature of the final state kinematics, the monochromatic result is supported on a $\delta$-function in momentum space, which complicates the comparison.   However, we can refine the monochromatic calculation in order to generate a result against which to benchmark, by including in the calculation a wavepacket for the initial photon. This is a natural physical refinement, {and also one which is suggested by the extra {volume factor $V^{-1}$} in the scattering probability (\ref{Pallm}), which can be removed by integration against a suitable density.} {The wavepacket essentially provides, here, additional integrals which render the singular monochromatic result finite, so that we can compare it with the LCFA, which must, of course, include the same wavepacket.} Including a photon wavepacket $\rho$ from the outset of the calculation corresponds to introducing into the S-matrix element (\ref{smatrix}) the factor,
\be \label{wp}
\int
\!
\frac{\ud l_- \ud^2 \bm{l}_\perp}{\sqrt{(2\pi)^3 2l_-}}
\rho(l_-,\bm{l}_\perp)
\ee
and carrying through the rest of the calculation as before.  This makes the S-matrix element, and subsequently the probability and cross section, formally very similar to that of nonlinear Compton scattering.

The simplest comparison is afforded by considering a wavepacket which is very broad and flat in momentum space, as then we can simply replace $\rho(l_-, \bm{l}_\perp) = 1$ and use the integrals in (\ref{wp}) to remove the $\delta$-function in the exact monochromatic result. More realistic wavepackets can be used, but this choice allows us to make analytic progress without requiring additional approximations which could lead to ambiguities in the comparison.

For the comparison between the LCFA and the exact calculation, we choose a circularly polarised monochromatic plane wave $a_\mu(\phi) = m a_0 (0,\cos\phi,\sin\phi,0)$ with dimensionless intensity parameter $a_0$. Using this along with the flat wavepacket we can express the absorption probability in terms of a sum over harmonics~\cite{RitusReview,landau4}. The exact result is
\be
\mathbb{P}_{\text{MONO}}
=
\frac{2\pi^3 m^2\alpha}{k.p}
\frac{T_{LF}}{V_{LF}}
\sum_{n=1}^{\infty}
\int\limits_{0}^{\min(u_n,1)} 
\frac{\ud u}{(1 - u)^2}
\bigg\{
-
J_n^2(z_n)
+
\frac{1}{2}
a_0^2
F(u)
\left(
-
2
J_n^2(z_n)
+
J_{n+1}^2(z_n)
+
J_{n-1}^2(z_n)
\right)
\bigg\}
\label{exactmono}
,
\ee
where $J_n$ is the order $n$ Bessel function of the first kind and
\begin{align}
z_n
=&
\frac{
	2
	n
	a_0
}{\sqrt{1 + a_0^2}}
\sqrt{\frac{u}{u_n}
	\bigg(
	1
	-
	\frac{u}{u_n}
	\bigg)
}
,
\quad&
u_n =& \frac{2 n k.p}{m^2(1 + a_0^2)} \;,
\quad&
u
=&
\frac{k.l}{k.l + k.p}
,
\quad&
F(u)
=
1
+
\frac12\frac{u^2}{(1-u)}
.
\label{zm}
\end{align}
The upper bound on the integration over the variable $u$ in (\ref{exactmono}) is determined by momentum conservation, but can be seen imediately from the requirement $z_n \in \mathbb{R}$ and the definition of $u$. Note that we use $u$ here, rather than $s$ as above, for convenience. The corresponding LCFA result in a flat wavepacket is
\be
\mathbb{P}_{\text{LCFA}}
=
-
\frac{2\pi^3 \alpha m^2}{k.p}
\frac{T_{LF}}{V_{LF}}
\int_0^1
\frac{\ud u}{(1 - u)^2}
\left\{
\Ai_1(\hat{z})
+
\left(
\frac{2}{\hat{z}}
+
\chi_\gamma
\sqrt{\hat{z}}
\right)
\Ai^\prime(\hat{z})
\right\}
,
\label{lcfamono}
\ee
where
\be\label{zlcf}
\hat{z}
=
\left(
\frac{u}{\chi_e}
\right)^{2/3} \;, 
\qquad
\Ai_1(\hat{z})
:=
\int_{{\hat{z}}}^{\infty}
\!\ud x
\,
\Ai(x) \;,
\ee
and $\chi_e = a_0 k.p/m^2$, $\chi_\gamma = s \chi_e = u \chi_e /(1-u)$. Note that $\chi_e$ is a constant for this case.

The monochromatic results and their locally constant field approximations are shown in Fig.~\ref{FIG:MONO1} and Fig.~\ref{FIG:MONO2} for a field intensity $a_0 = 10$ and various values of the invariant {$b_0 = k.p/m^2$}. Also shown is the relative error in each case. The following broad results hold. The LCFA agrees best for larger values of $s$, the photon's lightfront momentum. Just as for nonlinear Compton scattering, the LCFA does not correctly reproduce the harmonic structure of the monochromatic case~\cite{Harvey:2014qla}. In general, away from the distinct harmonic structure, there is excellent agreement between the LCFA and the exact monochromatic result for high $a_0$, as should be expected, and \text{lower} {$b_0$}, as is also required for the validity of the LCFA~\cite{BaierA0B0,Khok,Dinu:2015aci}. {Thus, for the angularly integrated process we find that the regimes of validity of the LCFA match with our expectations based on other strong-field processes.}

{In Figs. \ref{FIG:MONO1} and \ref{FIG:MONO2} one notices some qualitative differences between the curve with the lowest value of electron energy invariant {$b_{0}$} and the others. In both the probability and spectrum there is a ``dip'' as respectively $u$ and $s$ are increased. This is a $\chi_e < 1$ effect, which is most readily explained by considering the argument $\hat z$ of the Airy function in the LCFA (see (\ref{lcfamono}) and (\ref{zlcf})) and focusing on the differential energy spectrum Fig.~\ref{FIG:MONO2}. {For a non-negligible contribution, the argument of the Airy function should be small; when it is large, the function is exponentially suppressed.} In terms of $s$, the argument of the Airy function is $\hat{z} = \big[s /\chi_e(1 + s)\big]^{2/3}$. For each other curve in Fig. \ref{FIG:MONO2} $\chi_e \geq 1$, and so the argument of the Airy function is relatively small. Thus, since the spectrum scales (almost) linearly with $s$, there is a general increase as $s$ becomes large. However, for $\chi_e < 1$, in the range $\chi_e < s \ll 1$ the argument of the Airy function becomes large due to the $\chi_e^{-2/3}$ scaling of $\hat{z}$, such that the spectrum is suppressed. Moving into high values of $s \gg 1$, then $\hat{z} \rightarrow \chi_e^{-2/3}$, and the spectrum begins to linearly increase again with $s$, as $\text{Ai}(\hat{z}) = \text{constant}$.}

%
\subsection{Photon absorption in a counterpropagating pulse}
%

%
%
{
	In the above we focussed on a comparison of the exact (\ref{exactmono}) and LCFA (\ref{lcfamono}) probability, and found agreement within the expected regimes of validity for the LCFA.
	However, numerical implementation of strong field QED processes in PIC codes usually goes via probability ``rates'', $\ud \mathbb{P}/\ud \phi$, which are used to determine whether or not quantum process occur at each time step. It is not obvious that such an interpretation is sound; it is not natural given that the QED calculation is asymptotic. This poses some some interesting questions about the use of LCFA rates in numerical simulations, which will be investigated further elsewhere \cite{Us3}.	
}
To demonstrate the properties of the cross section (in the LCFA) it is instructive to consider a particular description of the laser pulse. As such, the gauge potential is chosen to be a linearly polarised plane wave $a_\mu(\phi) = ma_{0}f(\phi)(0,1,0,0)$, and $f(\phi)$ the pulse profile
\begin{align}\label{eq:Pulse}
f(\phi) = \cos^2\left(\frac{\phi}{2N}\right)\sin\phi \;.
\end{align}
{The phase interval of the pulse is $-N\pi < \phi < N\pi$, where} $N$ is the number of cycles, such that for a given $N$ the pulse length is $T_{LF} = 2 \pi N$. We take the laser to propagate in the $-z$ direction, i.e. $k_\mu = \omega_0(1,0,0,1)$ with central frequency $\omega_0 = 1.55$~eV ($ 800$~nm wavelength). For simplicity we take the momenta of the incoming electron and photon to lie in the $x$-$z$ plane. The electron counterpropagates with respect to the laser, whereas the photon will be given a general offset angle with respect to both the electron and laser. Hence the electron momentum is $p_\mu = \gamma m (1,0,0,-\beta)$, with $\beta = \sqrt{1 - 1/\gamma^2}$ for a Lorentz factor $\gamma$, while the photon momentum is $l_\mu = \omega (1,\sin\vartheta,0,-\cos\vartheta)$, for energy $\omega$ and angle $\vartheta$ relative to the electron propagation direction. With these conventions the cross section depends explicitly on four parameters, $a_0$, $\gamma$, $\omega$, and $\vartheta$.

The cross section is dominated by the Airy function in (\ref{crossLCFA}). Its argument $\bar{z}(\phi)$, recall  (\ref{zbarz}),  obeys $\bar{z}(\phi) \ge 0$, and even for moderately large values, e.g. $\bar{z}(\phi) \gtrsim 3$, the Airy function is well approximated by its exponentially vanishing asymptotic limit, $\Ai(\bar{z}) \sim \exp[-(2/3)\bar{z}^{2/3}]$.  Hence it is only when $\bar{z}(\phi) \rightarrow 0$, such that the integrand in (\ref{crossLCFA}) becomes large, that one can obtain significant contributions to the cross section. This is also the case for e.g.~nonlinear Compton scattering, where significant contributions come from the local maxima of the electric field, since the argument of the Airy function is proportional to $\chi_e^{-2/3}(\phi)$.  There is a more subtle dependence on the phase $\phi$ in the case of absorption, though, as the Airy arguments also depends on the local momenta $\pi_p(\phi)$. This leads to an interplay between the potential $a_\mu(\phi)$ and the electromagnetic field $\sim a'_\mu(\phi)$ as we integrate over $\phi$ in the cross section. Peaks in the integrand occur at or near the zeros of the potential.  At these points, $\bar{z}(\phi)$ reaches its minimum, which is a consequence of the fact that as $a_\mu(\phi)\rightarrow 0$ then $l . \pi_p(\phi) \rightarrow l . p \sim 0$ for high energy electrons (i.e.~$\gamma \gg 1$). Away from the zeros of the gauge potential the argument $\bar{z}(\phi)$ becomes large, diverging at the zeros of the field $|a'(\phi)|$. (Hence, when the field switches off, the entire process is correctly suppressed, as it cannot occur without the background.)

\begin{figure}[h!]
	\includegraphics[width=0.48\textwidth]{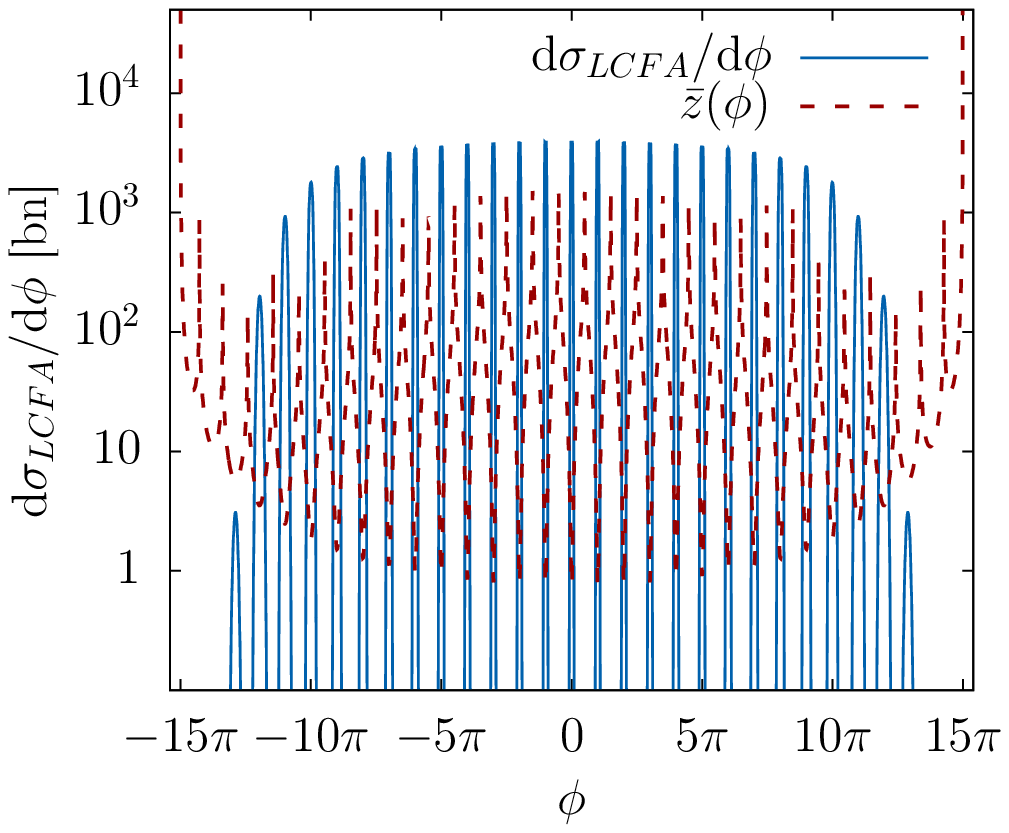}
	\includegraphics[width=0.48\textwidth]{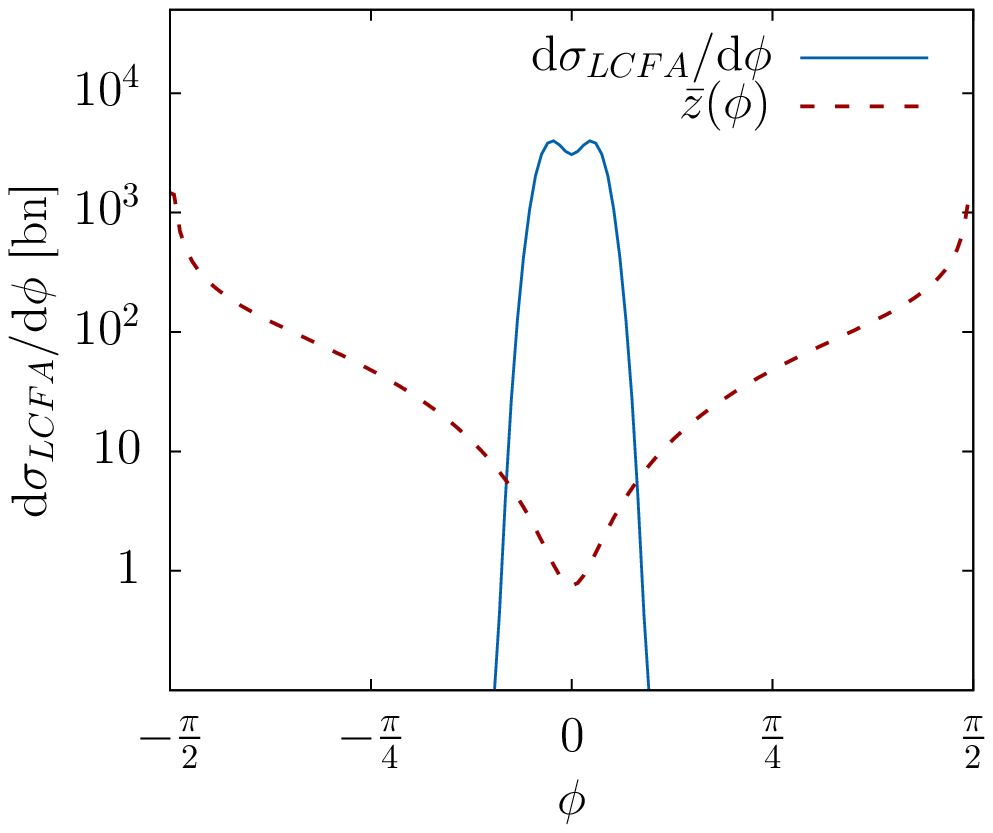}
	\caption{Differential cross section $\ud \sigma_{LCFA} / \ud \phi$ (solid blue line) with $a_0 = 10$, $\gamma = 500$, $\omega = 10m$ $\vartheta = 0$, for {an $N = 15$ cycle pulse ($\sim 40$~fs duration):} cross section over entire pulse duration (left) and central peak (right). Also shown is the argument of the Airy function $\bar{z}(\phi)$ (\ref{zbarz}) (red, dashed). The peaks of the differential cross section are located at the minima of $\bar{z}(\phi)$. For $\vartheta = 0$, these minima are where $a_\mu(\phi) \rightarrow 0$, as at these points $l.\pi_p(\phi) \rightarrow l.p \sim 0$ for high electron $\gamma$. {The parameters used correspond to $b_0 = k.p/m^2 = 0.003$ and $s = 0.02$, i.e. within the parameter region where the LCFA is in good agreement with the exact monochromatic calculation (see Fig \ref{FIG:MONO1} and \ref{FIG:MONO2}).}
	}
	\label{FIG:integrand}
\end{figure}

\begin{figure}[t!]
	\includegraphics[width=0.50\textwidth]{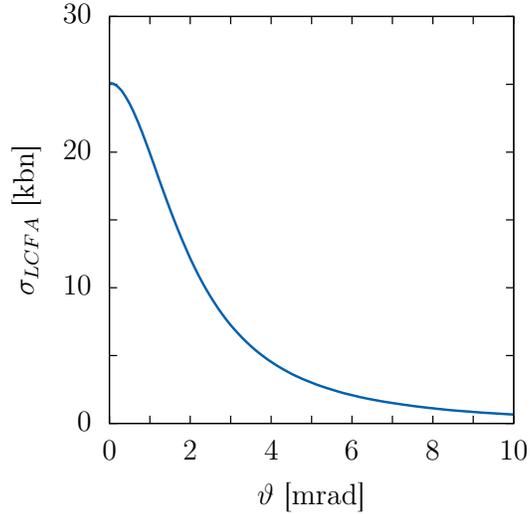}
	\caption{
		Dependence of the cross section $\sigma_{LCFA}$ on the photon angle $\vartheta$ for $a_0 = 10$,  $\gamma = 500$ and $\omega = 10m$, {such that $b_0 = 0.003$ and $s = 0.02$}.
		\label{FIG:CrossAngle}}
\end{figure}

Fig. \ref{FIG:integrand} illustrates this behaviour for $\gamma = 500$, photon energy $\omega = 10 m$ and background intensity $a_0 = 10$. We plot both the cross section (\ref{crossLCFA}) as a function of $\phi$ (solid blue line) as well as $\bar{z}(\phi)$ (dashed red line). Both the electron and photon are counterpropagating with respect to the laser direction ($ \vartheta = 0$), the pulse profile is as in (\ref{eq:Pulse}), with $N=15$ {corresponding to a pulse duration of $\sim 40$~fs}. The pulse profile (\ref{eq:Pulse}) has zeros at $\phi = \pm n \pi$ with $n \le N$, $n \in \mathbb{Z}$, and as discussed above, from Fig.~\ref{FIG:integrand} we see that each of the peaks in the integrand occur at these zeros; this is also where $\bar{z}(\phi)$ takes its minimum value.
Integrating over the laser phase $\phi$, the total cross section is found to be {$\sigma_{LCFA} \sim 2.5\times10^4$~bn}. At first glance this would appear to be an exceptionally likely process to occur, with the cross section being several orders of magnitude greater than that of, for example, {Thomson} scattering of an electron ($\sigma_T = 665.25$~mbn). However, this is primarily a consequence of the idealised geometry which was chosen, i.e. it is an artifact of the angle between the electron and photon propagation directions, $\vartheta = 0$. As the angle $\vartheta$ increases, there is a dramatic decay of the total integrated cross section $\sigma_{LCFA}$, see Fig. \ref{FIG:CrossAngle}. For the parameters used in Fig. \ref{FIG:integrand}, when the photon propagation direction is just a $\vartheta = 10$~mrad difference from the electron propagation direction, the cross section drops by around two orders of magnitude. When $\vartheta \ne 0$, the symmetry seen in the integrand for $\pm |\phi|$ also disappears.

{A closer inspection of the central peaks in the cross section in Fig. \ref{FIG:integrand} reveals a double peak structure, which is symmetric for the peak centred at $\phi = 0$ (shown in the right hand panel of the figure) and asymmetric for $|\phi| >0$. This is due to the relative contributions from the Airy function and pre-Airy terms in (\ref{crossLCFA}). The Airy function gives a series of symmetric peaks with maxima at $a_\mu =0$. However, the pre-Airy term achieves its \textit{minima} at $a_\mu=0$ for the central peak at $\phi=0$, and slightly offset for $|\phi| > 0$. The pre-Airy term scales with the electromagnetic field through electron $\chi_e$ as $\chi_e^{-2/3}(\phi)$, such that as we move away from the zeroes of the potential, we approach the zeroes of the field, which causes the pre-Airy term to blow up. Taking the product with the Airy function, this causes the double peak structure. Though the pre-Airy term diverges, this is not an immediate concern, as far enough away from the zeroes of the gauge potential, the Airy function suppresses the total cross section due to $\bar{z}$ becoming large.}

\subsection{Pulse shape effects and boundary contributions}\label{SECTION:WINGS}
%
%
\begin{figure}[t!]
	\includegraphics[width=0.48\textwidth]{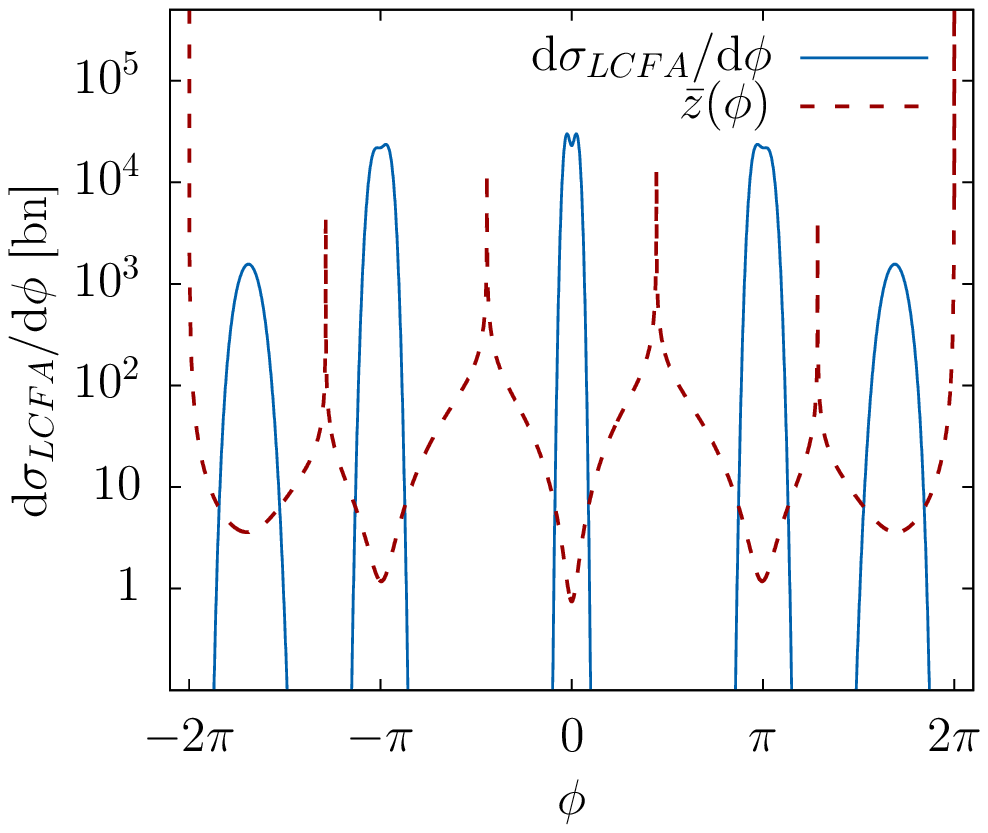}
	\includegraphics[width=0.4\textwidth,trim=1.5cm 0 0 0, clip]{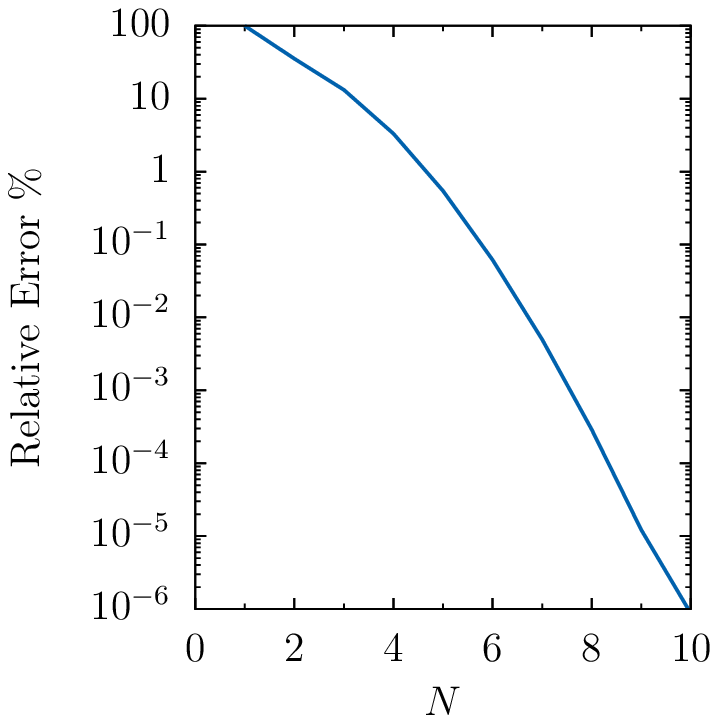}
	\caption{
		\label{FIG:integrandshort}
		\textit{Left:} {Differential cross section $\ud \sigma_{LCFA} / \ud \phi$ (solid blue line) with $a_0 = 10$, $\gamma = 500$, $\omega = 10m$ $\vartheta = 0$, {(i.e. $b_0 = 0.003$, $s = 0.02$)} for $N = 2$ corresponding to an ultra-short duration of $\sim 6$~fs. Also shown is the argument of the Airy function $\bar{z}(\phi)$ (\ref{zbarz}) (red, dashed).  \textit{Right}: The ``relative error'' incurred by excluding the initial and final half-cycle of the pulse, as a function of pulse length (number of cycles, $N$). Parameters as in Fig. \ref{FIG:integrand}.
	}}
\end{figure}

{So far we have considered the particular pulse profile (\ref{eq:Pulse}), and focussed on the experimentally realisable pulse duration of { $\sim 40$~fs}. The cross section will in general depend on the pulse shape and length, and in light of experimental efforts to produce ever shorter high-intensity laser pulses it is natural to consider the self-consistency of the absorption LCFA (\ref{crossLCFA}) for different pulse lengths and profiles. 
	
	The left hand panel of Fig. \ref{FIG:integrandshort} shows the differential cross section in an ultra-short pulse (\ref{eq:Pulse}) with $N=2$ {($\sim 5$~fs)}}. One sees surprisingly large contributions from the wings of the pulse where the field strength is low. These contributions, {we have checked,} can even be the dominant contribution to $\sigma$. The same behaviour is seen in the case of asymptotically switched pulses with e.g.~$\text{sech}^2$ or Gaussian envelopes, with large contributions to the cross section coming from up to $6$  standard deviations away from the peak of the pulse. This is problematic since, physically, there should not be contributions from regions of low field strength, when it is the presence of the field which allows the process to occur. The LCFA is of course not valid for low $a_0$, i.e.~in the wings of the pulse, but the situation for other processes is that the LCFA ``self-regulates'', with its typical Airy functions vanishing exponentially quickly in regions where the field strength goes to zero~\cite{Ilderton:2018nws}.

For both finite duration and asymptotically switched pulses, the origin of this unphysical behaviour can be identified in the LCFA cross section (\ref{crossLCFA}). Under the integral, the term with $4g\bar{z}$ is ``well behaved'' in the sense that whenever $\bar{z}$ becomes large, the Airy function suppresses the contribution, as we would expect from the consideration of other processes. The second term in the integrand, with a factor $z$, behaves differently. The parameter $z$ scales with the electromagnetic field as $\sim |a'(\phi)|^{-2/3}$, so becomes large in regions where the field goes to zero.  In the ``bulk'' of the pulse, when the field goes to zero the potential $a_\mu$ is at its maximum, and so $l.\pi_p$ becomes large (as it contains terms linear and quadratic in $a_\mu$, see (\ref{kinetic})). This makes $\bar{z}$ large, which causes the Airy function to again suppress contributions to $\sigma$. However, at the edges of the pulse, or far from the peak of the pulse in the case of asymptotically switched profiles, both the field \textit{and} the potential go to zero. In such regions, $\bar{z}$ does not become sufficiently large sufficiently quickly to suppress the large contribution coming from the pre-Airy term $z$, and this is what generates large contributions far from the centre of the pulse. {These unphysical contributions appear to also be a consequence of the particular geometry used, i.e. both the photon and electron counterpropagating with respect to the laser. As the angle $\vartheta$ is increased, the contributions from the edges of the pulse drastically reduce.} These results clearly point to a deficiency in the LCFA for short pulses, where there are high field gradients, and where there are long low-intensity ``tails''. Similar behaviour was seen in~\cite{King:2019cpj} in the LCFA for axion decay in a magnetic field of finite extent. The LCFA was shown to give a poor approximation for rapid turn on/off of the field.

We have found that for both finite duration and asymptotically switched pulses, these unphysical contributions become less significant at lower electron energy or higher incoming photon energy. In addition, for finite duration pulses such as (\ref{eq:Pulse}), the contribution to the cross section from the tails of the pulse can effectively be reduced by going to longer pulse duration. This can be seen by comparing the value obtained for the cross section (\ref{crossLCFA}) by integrating over the full pulse duration, $|\phi| < N \pi$, with the value obtained by excluding the initial and final half cycle of the potential, which for (\ref{eq:Pulse}) corresponds to integrating over the region $|\phi| < (N - 1) \pi$. This allows one to assess the relative contributions from the peaks at the beginning and end of the pulse. The right hand panel in Fig. \ref{FIG:integrandshort} shows that the relative contribution from the edges drops rapidly to zero as we go to pulse durations which are realisable with current and future laser technology.


\section{Conclusions}\label{sec:concs}
We have investigated the process of photon absorption by an electron in a background plane wave. The calculation of the scattering cross section initially follows standard methods in the Furry picture, but a closer investigation reveals a surprising depth of structure. We have shown that there is an {ambiguous boundary term in the amplitude with} a gauge-dependent coefficient. A simple regularisation of the integrals is enough to render the expression manifestly gauge {invariant} and unambiguous. However, {this results in a more complex} structure {at the level of the probability} than is typically considered in processes such as nonlinear Compton scattering and nonlinear Breit-Wheeler. The reason for the difference lies in the kinematics particular to ``2 to 1'' processes such as absorption; in nonlinear Compton, for example, {performing} the final state integrals can be used to affect the necessary regularisation as there are two outgoing particles, but for absorption this option is not available.

Perturbatively expanding the absorption S-matrix element for weak fields ($a_0 \ll 1$) demonstrates that the lowest-order contribution to the probability is not what one might na{\"i}vely expect for a process which is only possible in the {presence of the background (i.e ``laser-stimulated'')}. Momentum conservation requires that only the \textit{negative} frequency components of the background field contribute {in this perturbative limit}, which means that the perturbative expression is equivalent to absorbing the probe photon and emitting a photon \textit{into the laser field}. As such, a natural interpretation of the lowest-order process is as a contribution not to ``emissionless'' absorption of a photon by the electron (which is impossible in vacuum) but to the degenerate process of absorption \emph{with} photon emission, into an unobserved region of parameter space.    

Particle-in-cell codes, based on the locally constant field approximation (LCFA) provide a bridge between theory and experiment in strong-field QED. These codes currently neglect photon absorption (and the related process of pair annihilation). With an eye to future numerical implementation of absorption in such simulations, we have derived {its  LCFA}. This was compared with an exact analytical expression for absorption in a monochromatic plane wave (integrated over a suitable wavepacket in order to have a well-defined observable), and the expected regimes of validity were recovered. Again due to there being only a single particle in the out state, the LCFA for absorption has a very similar structure to the \emph{angularly resolved} LCFA for processes such as nonlinear Compton. In particular it depends not only on the quantum nonlinearity parameters ($\chi$) of the incoming particles, but also on the electron's instantaneous kinetic momentum. The cross section has a strong dependence on the angle between the incoming photon and electron, and also on the pulse profile and length. 

{The monochromatic result provides a useful benchmark not only because it is simple to calculate and interpret, but also because any experimental realisation of photon absorption will necessarily involve the overlap of electron, photon and laser species, for which a longer pulse duration is more suitable, and pulse-envelope effects become less important. (For example, the proposed LUXE experiment will use pulses of $\sim 35$~fs \cite{Abramowicz:2019gvx}, i.e.~order 10 cycles). It is nevertheless interesting to also consider ultra-short pulses. In this case,} we demonstrated that seemingly unphysical contributions occur near the edges of the pulse, but that these become negligible at longer (and more easily physically realisable) pulse lengths. This highlights a failure of the LCFA for ultra-short pulses, {which suggests that a more in-depth investigation of short pulse and wavepacket effects is motivated. These effects may become more important in absorption due to its particular kinematics; this will be pursued elsewhere. For an initial investigation of wavepacket effects see~\cite{Us3}.}

The depth of structure exhibited by the apparently simple absorption process demonstrates that investigations such as ours are necessary for a proper understanding of processes which are currently neglected by simulation schemes.

\begin{acknowledgments}
	We thank T.~Blackburn, M.~Marklund, {D.~Seipt} and S.~Tang for useful discussions. The authors are supported by the EPSRC, grant EP/S010319/1. 
\end{acknowledgments}


\begin{thebibliography}{56}%
	\makeatletter
	\providecommand \@ifxundefined [1]{%
		\@ifx{#1\undefined}
	}%
	\providecommand \@ifnum [1]{%
		\ifnum #1\expandafter \@firstoftwo
		\else \expandafter \@secondoftwo
		\fi
	}%
	\providecommand \@ifx [1]{%
		\ifx #1\expandafter \@firstoftwo
		\else \expandafter \@secondoftwo
		\fi
	}%
	\providecommand \natexlab [1]{#1}%
	\providecommand \enquote  [1]{``#1''}%
	\providecommand \bibnamefont  [1]{#1}%
	\providecommand \bibfnamefont [1]{#1}%
	\providecommand \citenamefont [1]{#1}%
	\providecommand \href@noop [0]{\@secondoftwo}%
	\providecommand \href [0]{\begingroup \@sanitize@url \@href}%
	\providecommand \@href[1]{\@@startlink{#1}\@@href}%
	\providecommand \@@href[1]{\endgroup#1\@@endlink}%
	\providecommand \@sanitize@url [0]{\catcode `\\12\catcode `\$12\catcode
		`\&12\catcode `\#12\catcode `\^12\catcode `\_12\catcode `\%12\relax}%
	\providecommand \@@startlink[1]{}%
	\providecommand \@@endlink[0]{}%
	\providecommand \url  [0]{\begingroup\@sanitize@url \@url }%
	\providecommand \@url [1]{\endgroup\@href {#1}{\urlprefix }}%
	\providecommand \urlprefix  [0]{URL }%
	\providecommand \Eprint [0]{\href }%
	\providecommand \doibase [0]{http://dx.doi.org/}%
	\providecommand \selectlanguage [0]{\@gobble}%
	\providecommand \bibinfo  [0]{\@secondoftwo}%
	\providecommand \bibfield  [0]{\@secondoftwo}%
	\providecommand \translation [1]{[#1]}%
	\providecommand \BibitemOpen [0]{}%
	\providecommand \bibitemStop [0]{}%
	\providecommand \bibitemNoStop [0]{.\EOS\space}%
	\providecommand \EOS [0]{\spacefactor3000\relax}%
	\providecommand \BibitemShut  [1]{\csname bibitem#1\endcsname}%
	\let\auto@bib@innerbib\@empty
	\bibitem [{\citenamefont {Abramowicz}\ \emph {et~al.}(2019)\citenamefont
		{Abramowicz} \emph {et~al.}}]{Abramowicz:2019gvx}%
	\BibitemOpen
	\bibfield  {author} {\bibinfo {author} {\bibfnamefont {H.}~\bibnamefont
			{Abramowicz}} \emph {et~al.},\ }\href@noop {} {\  (\bibinfo {year} {2019})},\
	\Eprint {http://arxiv.org/abs/1909.00860} {arXiv:1909.00860
		[physics.ins-det]} \BibitemShut {NoStop}%
	\bibitem [{\citenamefont {King}\ and\ \citenamefont
		{Heinzl}(2016)}]{King:2015tba}%
	\BibitemOpen
	\bibfield  {author} {\bibinfo {author} {\bibfnamefont {B.}~\bibnamefont
			{King}}\ and\ \bibinfo {author} {\bibfnamefont {T.}~\bibnamefont {Heinzl}},\
	}\href {\doibase 10.1017/hpl.2016.1} {\bibfield  {journal} {\bibinfo
			{journal} {High Power Laser Science and Engineering}\ }\textbf {\bibinfo
			{volume} {4}},\ \bibinfo {pages} {e5} (\bibinfo {year} {2016})},\ \Eprint
	{http://arxiv.org/abs/1510.08456} {arXiv:1510.08456 [hep-ph]} \BibitemShut
	{NoStop}%
	\bibitem [{\citenamefont {Fedotov}(2017)}]{Fedotov:2016afw}%
	\BibitemOpen
	\bibfield  {author} {\bibinfo {author} {\bibfnamefont {A.~M.}\ \bibnamefont
			{Fedotov}},\ }\bibfield  {booktitle} {\emph {\bibinfo {booktitle}
			{{Proceedings of the 25th International Laser Physics Workshop (LPHYS'16):
					Yerevan, Armenia, July 11-15, 2016}}},\ }\href {\doibase
		10.1088/1742-6596/826/1/012027} {\bibfield  {journal} {\bibinfo  {journal}
			{J. Phys. Conf. Ser.}\ }\textbf {\bibinfo {volume} {826}},\ \bibinfo {pages}
		{012027} (\bibinfo {year} {2017})},\ \Eprint
	{http://arxiv.org/abs/1608.02261} {arXiv:1608.02261 [hep-ph]} \BibitemShut
	{NoStop}%
	\bibitem [{\citenamefont {Yakimenko}\ \emph {et~al.}(2019)\citenamefont
		{Yakimenko} \emph {et~al.}}]{Yakimenko:2018kih}%
	\BibitemOpen
	\bibfield  {author} {\bibinfo {author} {\bibfnamefont {V.}~\bibnamefont
			{Yakimenko}} \emph {et~al.},\ }\href {\doibase
		10.1103/PhysRevLett.122.190404} {\bibfield  {journal} {\bibinfo  {journal}
			{Phys. Rev. Lett.}\ }\textbf {\bibinfo {volume} {122}},\ \bibinfo {pages}
		{190404} (\bibinfo {year} {2019})},\ \Eprint
	{http://arxiv.org/abs/1807.09271} {arXiv:1807.09271 [physics.plasm-ph]}
	\BibitemShut {NoStop}%
	\bibitem [{\citenamefont {Blackburn}\ \emph
		{et~al.}(2019{\natexlab{a}})\citenamefont {Blackburn}, \citenamefont
		{Ilderton}, \citenamefont {Marklund},\ and\ \citenamefont
		{Ridgers}}]{Blackburn:2018tsn}%
	\BibitemOpen
	\bibfield  {author} {\bibinfo {author} {\bibfnamefont {T.~G.}\ \bibnamefont
			{Blackburn}}, \bibinfo {author} {\bibfnamefont {A.}~\bibnamefont {Ilderton}},
		\bibinfo {author} {\bibfnamefont {M.}~\bibnamefont {Marklund}}, \ and\
		\bibinfo {author} {\bibfnamefont {C.~P.}\ \bibnamefont {Ridgers}},\ }\href
	{\doibase 10.1088/1367-2630/ab1e0d} {\bibfield  {journal} {\bibinfo
			{journal} {New J. Phys.}\ }\textbf {\bibinfo {volume} {21}},\ \bibinfo
		{pages} {053040} (\bibinfo {year} {2019}{\natexlab{a}})},\ \Eprint
	{http://arxiv.org/abs/1807.03730} {arXiv:1807.03730 [physics.plasm-ph]}
	\BibitemShut {NoStop}%
	\bibitem [{\citenamefont {Baumann}\ \emph {et~al.}(2019)\citenamefont
		{Baumann}, \citenamefont {Nerush}, \citenamefont {Pukhov},\ and\
		\citenamefont {Kostyukov}}]{Baumann:2018ovl}%
	\BibitemOpen
	\bibfield  {author} {\bibinfo {author} {\bibfnamefont {C.}~\bibnamefont
			{Baumann}}, \bibinfo {author} {\bibfnamefont {E.~N.}\ \bibnamefont {Nerush}},
		\bibinfo {author} {\bibfnamefont {A.}~\bibnamefont {Pukhov}}, \ and\ \bibinfo
		{author} {\bibfnamefont {I.~{\relax Yu}.}\ \bibnamefont {Kostyukov}},\ }\href
	{\doibase 10.1038/s41598-019-45582-5} {\bibfield  {journal} {\bibinfo
			{journal} {Sci. Rep.}\ }\textbf {\bibinfo {volume} {9}},\ \bibinfo {pages}
		{9407} (\bibinfo {year} {2019})},\ \Eprint {http://arxiv.org/abs/1811.03990}
	{arXiv:1811.03990 [physics.plasm-ph]} \BibitemShut {NoStop}%
	\bibitem [{\citenamefont {Harvey}\ \emph {et~al.}(2015)\citenamefont {Harvey},
		\citenamefont {Ilderton},\ and\ \citenamefont {King}}]{Harvey:2014qla}%
	\BibitemOpen
	\bibfield  {author} {\bibinfo {author} {\bibfnamefont {C.~N.}\ \bibnamefont
			{Harvey}}, \bibinfo {author} {\bibfnamefont {A.}~\bibnamefont {Ilderton}}, \
		and\ \bibinfo {author} {\bibfnamefont {B.}~\bibnamefont {King}},\ }\href
	{\doibase 10.1103/PhysRevA.91.013822} {\bibfield  {journal} {\bibinfo
			{journal} {Phys. Rev.}\ }\textbf {\bibinfo {volume} {A91}},\ \bibinfo {pages}
		{013822} (\bibinfo {year} {2015})},\ \Eprint {http://arxiv.org/abs/1409.6187}
	{arXiv:1409.6187 [physics.plasm-ph]} \BibitemShut {NoStop}%
	\bibitem [{\citenamefont {Di~Piazza}\ \emph {et~al.}(2018)\citenamefont
		{Di~Piazza}, \citenamefont {Tamburini}, \citenamefont {Meuren},\ and\
		\citenamefont {Keitel}}]{MeurenLCFA}%
	\BibitemOpen
	\bibfield  {author} {\bibinfo {author} {\bibfnamefont {A.}~\bibnamefont
			{Di~Piazza}}, \bibinfo {author} {\bibfnamefont {M.}~\bibnamefont
			{Tamburini}}, \bibinfo {author} {\bibfnamefont {S.}~\bibnamefont {Meuren}}, \
		and\ \bibinfo {author} {\bibfnamefont {C.~H.}\ \bibnamefont {Keitel}},\
	}\href {\doibase 10.1103/PhysRevA.98.012134} {\bibfield  {journal} {\bibinfo
			{journal} {Phys. Rev. A}\ }\textbf {\bibinfo {volume} {98}},\ \bibinfo
		{pages} {012134} (\bibinfo {year} {2018})}\BibitemShut {NoStop}%
	\bibitem [{\citenamefont {Ilderton}\ \emph
		{et~al.}(2019{\natexlab{a}})\citenamefont {Ilderton}, \citenamefont {King},\
		and\ \citenamefont {Seipt}}]{Ilderton:2018nws}%
	\BibitemOpen
	\bibfield  {author} {\bibinfo {author} {\bibfnamefont {A.}~\bibnamefont
			{Ilderton}}, \bibinfo {author} {\bibfnamefont {B.}~\bibnamefont {King}}, \
		and\ \bibinfo {author} {\bibfnamefont {D.}~\bibnamefont {Seipt}},\ }\href
	{\doibase 10.1103/PhysRevA.99.042121} {\bibfield  {journal} {\bibinfo
			{journal} {Phys. Rev.}\ }\textbf {\bibinfo {volume} {A99}},\ \bibinfo {pages}
		{042121} (\bibinfo {year} {2019}{\natexlab{a}})},\ \Eprint
	{http://arxiv.org/abs/1808.10339} {arXiv:1808.10339 [hep-ph]} \BibitemShut
	{NoStop}%
	\bibitem [{\citenamefont {Cole}\ \emph {et~al.}(2018)\citenamefont {Cole} \emph
		{et~al.}}]{Cole:2017zca}%
	\BibitemOpen
	\bibfield  {author} {\bibinfo {author} {\bibfnamefont {J.~M.}\ \bibnamefont
			{Cole}} \emph {et~al.},\ }\href {\doibase 10.1103/PhysRevX.8.011020}
	{\bibfield  {journal} {\bibinfo  {journal} {Phys. Rev.}\ }\textbf {\bibinfo
			{volume} {X8}},\ \bibinfo {pages} {011020} (\bibinfo {year} {2018})},\
	\Eprint {http://arxiv.org/abs/1707.06821} {arXiv:1707.06821
		[physics.plasm-ph]} \BibitemShut {NoStop}%
	\bibitem [{\citenamefont {Poder}\ \emph {et~al.}(2018)\citenamefont {Poder}
		\emph {et~al.}}]{Poder:2018ifi}%
	\BibitemOpen
	\bibfield  {author} {\bibinfo {author} {\bibfnamefont {K.}~\bibnamefont
			{Poder}} \emph {et~al.},\ }\href {\doibase 10.1103/PhysRevX.8.031004}
	{\bibfield  {journal} {\bibinfo  {journal} {Phys. Rev.}\ }\textbf {\bibinfo
			{volume} {X8}},\ \bibinfo {pages} {031004} (\bibinfo {year} {2018})},\
	\Eprint {http://arxiv.org/abs/1709.01861} {arXiv:1709.01861
		[physics.plasm-ph]} \BibitemShut {NoStop}%
	\bibitem [{\citenamefont {Zhukovskii}\ and\ \citenamefont
		{Nikitina}(1973)}]{zhukovskii73}%
	\BibitemOpen
	\bibfield  {author} {\bibinfo {author} {\bibfnamefont {V.~C.}\ \bibnamefont
			{Zhukovskii}}\ and\ \bibinfo {author} {\bibfnamefont {N.~S.}\ \bibnamefont
			{Nikitina}},\ }\href@noop {} {\bibfield  {journal} {\bibinfo  {journal} {{Zh.
					Eksp. Teor. Fiz.}}\ }\textbf {\bibinfo {volume} {64}},\ \bibinfo {pages}
		{1169} (\bibinfo {year} {1973})}\BibitemShut {NoStop}%
	\bibitem [{\citenamefont {Di~Piazza}\ \emph {et~al.}(2008)\citenamefont
		{Di~Piazza}, \citenamefont {Hatsagortsyan},\ and\ \citenamefont
		{Keitel}}]{dipiazza08b}%
	\BibitemOpen
	\bibfield  {author} {\bibinfo {author} {\bibfnamefont {A.}~\bibnamefont
			{Di~Piazza}}, \bibinfo {author} {\bibfnamefont {K.~Z.}\ \bibnamefont
			{Hatsagortsyan}}, \ and\ \bibinfo {author} {\bibfnamefont {C.~H.}\
			\bibnamefont {Keitel}},\ }\href@noop {} {\bibfield  {journal} {\bibinfo
			{journal} {Phys. Rev. A}\ }\textbf {\bibinfo {volume} {78}},\ \bibinfo
		{pages} {062109} (\bibinfo {year} {2008})}\BibitemShut {NoStop}%
	\bibitem [{\citenamefont {Hu}\ and\ \citenamefont
		{M{\"u}ller}(2011)}]{Hu:2011eq}%
	\BibitemOpen
	\bibfield  {author} {\bibinfo {author} {\bibfnamefont {H.}~\bibnamefont
			{Hu}}\ and\ \bibinfo {author} {\bibfnamefont {C.}~\bibnamefont
			{M{\"u}ller}},\ }\href {\doibase 10.1103/PhysRevLett.107.090402} {\bibfield
		{journal} {\bibinfo  {journal} {Phys. Rev. Lett.}\ }\textbf {\bibinfo
			{volume} {107}},\ \bibinfo {pages} {090402} (\bibinfo {year} {2011})},\
	\Eprint {http://arxiv.org/abs/1105.0279} {arXiv:1105.0279 [hep-ph]}
	\BibitemShut {NoStop}%
	\bibitem [{\citenamefont {Gies}\ \emph {et~al.}(2014)\citenamefont {Gies},
		\citenamefont {Karbstein},\ and\ \citenamefont {Shaisultanov}}]{gies14}%
	\BibitemOpen
	\bibfield  {author} {\bibinfo {author} {\bibfnamefont {H.}~\bibnamefont
			{Gies}}, \bibinfo {author} {\bibfnamefont {F.}~\bibnamefont {Karbstein}}, \
		and\ \bibinfo {author} {\bibfnamefont {R.}~\bibnamefont {Shaisultanov}},\
	}\href {\doibase 10.1103/PhysRevD.90.033007} {\bibfield  {journal} {\bibinfo
			{journal} {Phys. Rev. D}\ }\textbf {\bibinfo {volume} {90}},\ \bibinfo
		{pages} {033007} (\bibinfo {year} {2014})}\BibitemShut {NoStop}%
	\bibitem [{\citenamefont {B\"ohl}\ \emph {et~al.}(2015)\citenamefont {B\"ohl},
		\citenamefont {King},\ and\ \citenamefont {Ruhl}}]{king15c}%
	\BibitemOpen
	\bibfield  {author} {\bibinfo {author} {\bibfnamefont {P.}~\bibnamefont
			{B\"ohl}}, \bibinfo {author} {\bibfnamefont {B.}~\bibnamefont {King}}, \ and\
		\bibinfo {author} {\bibfnamefont {H.}~\bibnamefont {Ruhl}},\ }\href {\doibase
		10.1103/PhysRevA.92.032115} {\bibfield  {journal} {\bibinfo  {journal} {Phys.
				Rev. A}\ }\textbf {\bibinfo {volume} {92}},\ \bibinfo {pages} {032115}
		(\bibinfo {year} {2015})}\BibitemShut {NoStop}%
	\bibitem [{\citenamefont {Nedoreshta}\ \emph {et~al.}(2015)\citenamefont
		{Nedoreshta}, \citenamefont {Roshchupkin},\ and\ \citenamefont
		{Voroshilo}}]{Vorosh15}%
	\BibitemOpen
	\bibfield  {author} {\bibinfo {author} {\bibfnamefont {V.~N.}\ \bibnamefont
			{Nedoreshta}}, \bibinfo {author} {\bibfnamefont {S.~P.}\ \bibnamefont
			{Roshchupkin}}, \ and\ \bibinfo {author} {\bibfnamefont {A.~I.}\ \bibnamefont
			{Voroshilo}},\ }\href {\doibase 10.1103/PhysRevA.91.062110} {\bibfield
		{journal} {\bibinfo  {journal} {Phys. Rev. A}\ }\textbf {\bibinfo {volume}
			{91}},\ \bibinfo {pages} {062110} (\bibinfo {year} {2015})}\BibitemShut
	{NoStop}%
	\bibitem [{\citenamefont {Gies}\ \emph {et~al.}(2016)\citenamefont {Gies},
		\citenamefont {Karbstein},\ and\ \citenamefont {Seegert}}]{gies16}%
	\BibitemOpen
	\bibfield  {author} {\bibinfo {author} {\bibfnamefont {H.}~\bibnamefont
			{Gies}}, \bibinfo {author} {\bibfnamefont {F.}~\bibnamefont {Karbstein}}, \
		and\ \bibinfo {author} {\bibfnamefont {N.}~\bibnamefont {Seegert}},\ }\href
	{\doibase 10.1103/PhysRevD.93.085034} {\bibfield  {journal} {\bibinfo
			{journal} {Phys. Rev. D}\ }\textbf {\bibinfo {volume} {93}},\ \bibinfo
		{pages} {085034} (\bibinfo {year} {2016})}\BibitemShut {NoStop}%
	\bibitem [{\citenamefont {Hartin}(2006)}]{Hartin:2017uah}%
	\BibitemOpen
	\bibfield  {author} {\bibinfo {author} {\bibfnamefont {A.}~\bibnamefont
			{Hartin}},\ }\emph {\bibinfo {title} {{Second Order QED Processes in an
				Intense Electromagnetic Field}}},\ \href@noop {} {Ph.D. thesis},\ \bibinfo
	{school} {Queen Mary, U. of London} (\bibinfo {year} {2006}),\ \Eprint
	{http://arxiv.org/abs/1701.02906} {arXiv:1701.02906 [hep-ph]} \BibitemShut
	{NoStop}%
	\bibitem [{\citenamefont {Ritus}(1985)}]{RitusReview}%
	\BibitemOpen
	\bibfield  {author} {\bibinfo {author} {\bibfnamefont {V.~I.}\ \bibnamefont
			{Ritus}},\ }\href@noop {} {\bibfield  {journal} {\bibinfo  {journal} {J.
				Russ. Laser Res.}\ }\textbf {\bibinfo {volume} {6}},\ \bibinfo {pages} {497}
		(\bibinfo {year} {1985})}\BibitemShut {NoStop}%
	\bibitem [{\citenamefont {Gonoskov}\ \emph {et~al.}(2015)\citenamefont
		{Gonoskov}, \citenamefont {Bastrakov}, \citenamefont {Efimenko},
		\citenamefont {Ilderton}, \citenamefont {Marklund}, \citenamefont {Meyerov},
		\citenamefont {Muraviev}, \citenamefont {Sergeev}, \citenamefont {Surmin},\
		and\ \citenamefont {Wallin}}]{Gonoskov:2014mda}%
	\BibitemOpen
	\bibfield  {author} {\bibinfo {author} {\bibfnamefont {A.}~\bibnamefont
			{Gonoskov}}, \bibinfo {author} {\bibfnamefont {S.}~\bibnamefont {Bastrakov}},
		\bibinfo {author} {\bibfnamefont {E.}~\bibnamefont {Efimenko}}, \bibinfo
		{author} {\bibfnamefont {A.}~\bibnamefont {Ilderton}}, \bibinfo {author}
		{\bibfnamefont {M.}~\bibnamefont {Marklund}}, \bibinfo {author}
		{\bibfnamefont {I.}~\bibnamefont {Meyerov}}, \bibinfo {author} {\bibfnamefont
			{A.}~\bibnamefont {Muraviev}}, \bibinfo {author} {\bibfnamefont
			{A.}~\bibnamefont {Sergeev}}, \bibinfo {author} {\bibfnamefont
			{I.}~\bibnamefont {Surmin}}, \ and\ \bibinfo {author} {\bibfnamefont
			{E.}~\bibnamefont {Wallin}},\ }\href {\doibase 10.1103/PhysRevE.92.023305}
	{\bibfield  {journal} {\bibinfo  {journal} {Phys. Rev.}\ }\textbf {\bibinfo
			{volume} {E92}},\ \bibinfo {pages} {023305} (\bibinfo {year} {2015})},\
	\Eprint {http://arxiv.org/abs/1412.6426} {arXiv:1412.6426 [physics.plasm-ph]}
	\BibitemShut {NoStop}%
	\bibitem [{\citenamefont {Dinu}\ \emph {et~al.}(2012)\citenamefont {Dinu},
		\citenamefont {Heinzl},\ and\ \citenamefont {Ilderton}}]{Dinu:2012tj}%
	\BibitemOpen
	\bibfield  {author} {\bibinfo {author} {\bibfnamefont {V.}~\bibnamefont
			{Dinu}}, \bibinfo {author} {\bibfnamefont {T.}~\bibnamefont {Heinzl}}, \ and\
		\bibinfo {author} {\bibfnamefont {A.}~\bibnamefont {Ilderton}},\ }\href
	{\doibase 10.1103/PhysRevD.86.085037} {\bibfield  {journal} {\bibinfo
			{journal} {Phys. Rev.}\ }\textbf {\bibinfo {volume} {D86}},\ \bibinfo {pages}
		{085037} (\bibinfo {year} {2012})},\ \Eprint {http://arxiv.org/abs/1206.3957}
	{arXiv:1206.3957 [hep-ph]} \BibitemShut {NoStop}%
	\bibitem [{\citenamefont {Volkov}(1935)}]{volkov35}%
	\BibitemOpen
	\bibfield  {author} {\bibinfo {author} {\bibfnamefont {D.~M.}\ \bibnamefont
			{Volkov}},\ }\href@noop {} {\ \textbf {\bibinfo {volume} {94}},\ \bibinfo
		{pages} {250} (\bibinfo {year} {1935})}\BibitemShut {NoStop}%
	\bibitem [{\citenamefont {Furry}(1951)}]{Furry51}%
	\BibitemOpen
	\bibfield  {author} {\bibinfo {author} {\bibfnamefont {W.~H.}\ \bibnamefont
			{Furry}},\ }\href@noop {} {\ \textbf {\bibinfo {volume} {81}},\ \bibinfo
		{pages} {115} (\bibinfo {year} {1951})}\BibitemShut {NoStop}%
	\bibitem [{\citenamefont {Seipt}(2017)}]{Seipt:2017ckc}%
	\BibitemOpen
	\bibfield  {author} {\bibinfo {author} {\bibfnamefont {D.}~\bibnamefont
			{Seipt}},\ }in\ \href {\doibase 10.3204/DESY-PROC-2016-04/Seipt} {\emph
		{\bibinfo {booktitle} {{Proceedings, Quantum Field Theory at the Limits: from
					Strong Fields to Heavy Quarks (HQ 2016): Dubna, Russia, July 18-30, 2016}}}}\
	(\bibinfo {year} {2017})\ pp.\ \bibinfo {pages} {24--43},\ \Eprint
	{http://arxiv.org/abs/1701.03692} {arXiv:1701.03692 [physics.plasm-ph]}
	\BibitemShut {NoStop}%
	\bibitem [{\citenamefont {Peskin}\ and\ \citenamefont
		{Schroeder}(1995)}]{Peskin:1995ev}%
	\BibitemOpen
	\bibfield  {author} {\bibinfo {author} {\bibfnamefont {M.~E.}\ \bibnamefont
			{Peskin}}\ and\ \bibinfo {author} {\bibfnamefont {D.~V.}\ \bibnamefont
			{Schroeder}},\ }\href {http://www.slac.stanford.edu/~mpeskin/QFT.html} {\emph
		{\bibinfo {title} {{An Introduction to quantum field theory}}}}\ (\bibinfo
	{publisher} {Addison-Wesley},\ \bibinfo {address} {Reading, USA},\ \bibinfo
	{year} {1995})\BibitemShut {NoStop}%
	\bibitem [{\citenamefont {Ilderton}\ and\ \citenamefont
		{Torgrimsson}(2013)}]{Ilderton:2012qe}%
	\BibitemOpen
	\bibfield  {author} {\bibinfo {author} {\bibfnamefont {A.}~\bibnamefont
			{Ilderton}}\ and\ \bibinfo {author} {\bibfnamefont {G.}~\bibnamefont
			{Torgrimsson}},\ }\href {\doibase 10.1103/PhysRevD.87.085040} {\bibfield
		{journal} {\bibinfo  {journal} {Phys. Rev.}\ }\textbf {\bibinfo {volume}
			{D87}},\ \bibinfo {pages} {085040} (\bibinfo {year} {2013})},\ \Eprint
	{http://arxiv.org/abs/1210.6840} {arXiv:1210.6840 [hep-th]} \BibitemShut
	{NoStop}%
	\bibitem [{\citenamefont {Ilderton}\ \emph
		{et~al.}(2019{\natexlab{b}})\citenamefont {Ilderton}, \citenamefont {King},\
		and\ \citenamefont {Tang}}]{Us3}%
	\BibitemOpen
	\bibfield  {author} {\bibinfo {author} {\bibfnamefont {A.}~\bibnamefont
			{Ilderton}}, \bibinfo {author} {\bibfnamefont {B.}~\bibnamefont {King}}, \
		and\ \bibinfo {author} {\bibfnamefont {S.}~\bibnamefont {Tang}},\ }\href@noop
	{} {\  (\bibinfo {year} {2019}{\natexlab{b}})},\ \Eprint
	{http://arxiv.org/abs/1909.01141} {arXiv:1909.01141 [physics.plasm-ph]}
	\BibitemShut {NoStop}%
	\bibitem [{\citenamefont {Boca}\ and\ \citenamefont
		{Florescu}(2009)}]{Boca:2009zz}%
	\BibitemOpen
	\bibfield  {author} {\bibinfo {author} {\bibfnamefont {M.}~\bibnamefont
			{Boca}}\ and\ \bibinfo {author} {\bibfnamefont {V.}~\bibnamefont
			{Florescu}},\ }\href {\doibase 10.1103/PhysRevA.80.053403} {\bibfield
		{journal} {\bibinfo  {journal} {Phys. Rev.}\ }\textbf {\bibinfo {volume}
			{A80}},\ \bibinfo {pages} {053403} (\bibinfo {year} {2009})}\BibitemShut
	{NoStop}%
	\bibitem [{\citenamefont {Ilderton}(2011)}]{Ilderton:2010wr}%
	\BibitemOpen
	\bibfield  {author} {\bibinfo {author} {\bibfnamefont {A.}~\bibnamefont
			{Ilderton}},\ }\href {\doibase 10.1103/PhysRevLett.106.020404} {\bibfield
		{journal} {\bibinfo  {journal} {Phys. Rev. Lett.}\ }\textbf {\bibinfo
			{volume} {106}},\ \bibinfo {pages} {020404} (\bibinfo {year} {2011})},\
	\Eprint {http://arxiv.org/abs/1011.4072} {arXiv:1011.4072 [hep-ph]}
	\BibitemShut {NoStop}%
	\bibitem [{\citenamefont {Seipt}\ and\ \citenamefont
		{Kampfer}(2011)}]{Seipt:2010ya}%
	\BibitemOpen
	\bibfield  {author} {\bibinfo {author} {\bibfnamefont {D.}~\bibnamefont
			{Seipt}}\ and\ \bibinfo {author} {\bibfnamefont {B.}~\bibnamefont
			{Kampfer}},\ }\href {\doibase 10.1103/PhysRevA.83.022101} {\bibfield
		{journal} {\bibinfo  {journal} {Phys. Rev.}\ }\textbf {\bibinfo {volume}
			{A83}},\ \bibinfo {pages} {022101} (\bibinfo {year} {2011})},\ \Eprint
	{http://arxiv.org/abs/1010.3301} {arXiv:1010.3301 [hep-ph]} \BibitemShut
	{NoStop}%
	\bibitem [{\citenamefont {Mackenroth}\ and\ \citenamefont
		{Di~Piazza}(2011)}]{Mackenroth:2010jr}%
	\BibitemOpen
	\bibfield  {author} {\bibinfo {author} {\bibfnamefont {F.}~\bibnamefont
			{Mackenroth}}\ and\ \bibinfo {author} {\bibfnamefont {A.}~\bibnamefont
			{Di~Piazza}},\ }\href {\doibase 10.1103/PhysRevA.83.032106} {\bibfield
		{journal} {\bibinfo  {journal} {Phys. Rev.}\ }\textbf {\bibinfo {volume}
			{A83}},\ \bibinfo {pages} {032106} (\bibinfo {year} {2011})},\ \Eprint
	{http://arxiv.org/abs/1010.6251} {arXiv:1010.6251 [hep-ph]} \BibitemShut
	{NoStop}%
	\bibitem [{\citenamefont {Mackenroth}\ and\ \citenamefont
		{Di~Piazza}(2018)}]{Mackenroth:2018smh}%
	\BibitemOpen
	\bibfield  {author} {\bibinfo {author} {\bibfnamefont {F.}~\bibnamefont
			{Mackenroth}}\ and\ \bibinfo {author} {\bibfnamefont {A.}~\bibnamefont
			{Di~Piazza}},\ }\href {\doibase 10.1103/PhysRevD.98.116002} {\bibfield
		{journal} {\bibinfo  {journal} {Phys. Rev.}\ }\textbf {\bibinfo {volume}
			{D98}},\ \bibinfo {pages} {116002} (\bibinfo {year} {2018})},\ \Eprint
	{http://arxiv.org/abs/1805.01731} {arXiv:1805.01731 [hep-ph]} \BibitemShut
	{NoStop}%
	\bibitem [{\citenamefont {Dinu}\ and\ \citenamefont
		{Torgrimsson}(2018{\natexlab{a}})}]{Dinu:2017uoj}%
	\BibitemOpen
	\bibfield  {author} {\bibinfo {author} {\bibfnamefont {V.}~\bibnamefont
			{Dinu}}\ and\ \bibinfo {author} {\bibfnamefont {G.}~\bibnamefont
			{Torgrimsson}},\ }\href {\doibase 10.1103/PhysRevD.97.036021} {\bibfield
		{journal} {\bibinfo  {journal} {Phys. Rev.}\ }\textbf {\bibinfo {volume}
			{D97}},\ \bibinfo {pages} {036021} (\bibinfo {year} {2018}{\natexlab{a}})},\
	\Eprint {http://arxiv.org/abs/1711.04344} {arXiv:1711.04344 [hep-ph]}
	\BibitemShut {NoStop}%
	\bibitem [{\citenamefont {Dinu}\ and\ \citenamefont
		{Torgrimsson}(2018{\natexlab{b}})}]{Dinu:2018efz}%
	\BibitemOpen
	\bibfield  {author} {\bibinfo {author} {\bibfnamefont {V.}~\bibnamefont
			{Dinu}}\ and\ \bibinfo {author} {\bibfnamefont {G.}~\bibnamefont
			{Torgrimsson}},\ }\href@noop {} {\  (\bibinfo {year} {2018}{\natexlab{b}})},\
	\Eprint {http://arxiv.org/abs/1811.00451} {arXiv:1811.00451 [hep-ph]}
	\BibitemShut {NoStop}%
	\bibitem [{\citenamefont {Dinu}(2013)}]{Dinu:2013hsd}%
	\BibitemOpen
	\bibfield  {author} {\bibinfo {author} {\bibfnamefont {V.}~\bibnamefont
			{Dinu}},\ }\href {\doibase 10.1103/PhysRevA.87.052101} {\bibfield  {journal}
		{\bibinfo  {journal} {Phys. Rev.}\ }\textbf {\bibinfo {volume} {A87}},\
		\bibinfo {pages} {052101} (\bibinfo {year} {2013})},\ \Eprint
	{http://arxiv.org/abs/1302.1513} {arXiv:1302.1513 [hep-ph]} \BibitemShut
	{NoStop}%
	\bibitem [{\citenamefont {Yennie}\ \emph {et~al.}(1961)\citenamefont {Yennie},
		\citenamefont {Frautschi},\ and\ \citenamefont {Suura}}]{Yennie:1961ad}%
	\BibitemOpen
	\bibfield  {author} {\bibinfo {author} {\bibfnamefont {D.~R.}\ \bibnamefont
			{Yennie}}, \bibinfo {author} {\bibfnamefont {S.~C.}\ \bibnamefont
			{Frautschi}}, \ and\ \bibinfo {author} {\bibfnamefont {H.}~\bibnamefont
			{Suura}},\ }\href {\doibase 10.1016/0003-4916(61)90151-8} {\bibfield
		{journal} {\bibinfo  {journal} {Annals Phys.}\ }\textbf {\bibinfo {volume}
			{13}},\ \bibinfo {pages} {379} (\bibinfo {year} {1961})}\BibitemShut
	{NoStop}%
	\bibitem [{\citenamefont {Kibble}(1965)}]{Kibble:1965zza}%
	\BibitemOpen
	\bibfield  {author} {\bibinfo {author} {\bibfnamefont {T.~W.~B.}\
			\bibnamefont {Kibble}},\ }\href {\doibase 10.1103/PhysRev.138.B740}
	{\bibfield  {journal} {\bibinfo  {journal} {Phys. Rev.}\ }\textbf {\bibinfo
			{volume} {138}},\ \bibinfo {pages} {B740} (\bibinfo {year}
		{1965})}\BibitemShut {NoStop}%
	\bibitem [{\citenamefont {Frantz}(1965)}]{Frantz}%
	\BibitemOpen
	\bibfield  {author} {\bibinfo {author} {\bibfnamefont {L.~M.}\ \bibnamefont
			{Frantz}},\ }\href {\doibase 10.1103/PhysRev.139.B1326} {\bibfield  {journal}
		{\bibinfo  {journal} {Phys. Rev.}\ }\textbf {\bibinfo {volume} {139}},\
		\bibinfo {pages} {B1326} (\bibinfo {year} {1965})}\BibitemShut {NoStop}%
	\bibitem [{\citenamefont {Gavrilov}\ and\ \citenamefont
		{Gitman}(1990)}]{Gavrilov:1990qa}%
	\BibitemOpen
	\bibfield  {author} {\bibinfo {author} {\bibfnamefont {S.~P.}\ \bibnamefont
			{Gavrilov}}\ and\ \bibinfo {author} {\bibfnamefont {D.~M.}\ \bibnamefont
			{Gitman}},\ }\href@noop {} {\bibfield  {journal} {\bibinfo  {journal} {Sov.
				J. Nucl. Phys.}\ }\textbf {\bibinfo {volume} {51}},\ \bibinfo {pages} {1040}
		(\bibinfo {year} {1990})},\ \bibinfo {note} {[Yad.
		Fiz.51,1644(1990)]}\BibitemShut {NoStop}%
	\bibitem [{\citenamefont {Podszus}\ and\ \citenamefont
		{Di~Piazza}(2019)}]{Podszus:2018hnz}%
	\BibitemOpen
	\bibfield  {author} {\bibinfo {author} {\bibfnamefont {T.}~\bibnamefont
			{Podszus}}\ and\ \bibinfo {author} {\bibfnamefont {A.}~\bibnamefont
			{Di~Piazza}},\ }\href {\doibase 10.1103/PhysRevD.99.076004} {\bibfield
		{journal} {\bibinfo  {journal} {Phys. Rev.}\ }\textbf {\bibinfo {volume}
			{D99}},\ \bibinfo {pages} {076004} (\bibinfo {year} {2019})},\ \Eprint
	{http://arxiv.org/abs/1812.08673} {arXiv:1812.08673 [hep-ph]} \BibitemShut
	{NoStop}%
	\bibitem [{\citenamefont {Ilderton}(2019)}]{Ilderton:2019kqp}%
	\BibitemOpen
	\bibfield  {author} {\bibinfo {author} {\bibfnamefont {A.}~\bibnamefont
			{Ilderton}},\ }\href {\doibase 10.1103/PhysRevD.99.085002} {\bibfield
		{journal} {\bibinfo  {journal} {Phys. Rev.}\ }\textbf {\bibinfo {volume}
			{D99}},\ \bibinfo {pages} {085002} (\bibinfo {year} {2019})},\ \Eprint
	{http://arxiv.org/abs/1901.00317} {arXiv:1901.00317 [hep-ph]} \BibitemShut
	{NoStop}%
	\bibitem [{\citenamefont {King}\ \emph {et~al.}(2019)\citenamefont {King},
		\citenamefont {Dillon}, \citenamefont {Beyer},\ and\ \citenamefont
		{Gregori}}]{King:2019cpj}%
	\BibitemOpen
	\bibfield  {author} {\bibinfo {author} {\bibfnamefont {B.}~\bibnamefont
			{King}}, \bibinfo {author} {\bibfnamefont {B.~M.}\ \bibnamefont {Dillon}},
		\bibinfo {author} {\bibfnamefont {K.~A.}\ \bibnamefont {Beyer}}, \ and\
		\bibinfo {author} {\bibfnamefont {G.}~\bibnamefont {Gregori}},\ }\href@noop
	{} {\  (\bibinfo {year} {2019})},\ \Eprint {http://arxiv.org/abs/1905.05201}
	{arXiv:1905.05201 [hep-ph]} \BibitemShut {NoStop}%
	\bibitem [{\citenamefont {Di~Piazza}\ \emph {et~al.}(2019)\citenamefont
		{Di~Piazza}, \citenamefont {Tamburini}, \citenamefont {Meuren},\ and\
		\citenamefont {Keitel}}]{DiPiazza:2018bfu}%
	\BibitemOpen
	\bibfield  {author} {\bibinfo {author} {\bibfnamefont {A.}~\bibnamefont
			{Di~Piazza}}, \bibinfo {author} {\bibfnamefont {M.}~\bibnamefont
			{Tamburini}}, \bibinfo {author} {\bibfnamefont {S.}~\bibnamefont {Meuren}}, \
		and\ \bibinfo {author} {\bibfnamefont {C.~H.}\ \bibnamefont {Keitel}},\
	}\href {\doibase 10.1103/PhysRevA.99.022125} {\bibfield  {journal} {\bibinfo
			{journal} {Phys. Rev.}\ }\textbf {\bibinfo {volume} {A99}},\ \bibinfo {pages}
		{022125} (\bibinfo {year} {2019})},\ \Eprint
	{http://arxiv.org/abs/1811.05834} {arXiv:1811.05834 [hep-ph]} \BibitemShut
	{NoStop}%
	\bibitem [{\citenamefont {Blackburn}\ \emph
		{et~al.}(2019{\natexlab{b}})\citenamefont {Blackburn}, \citenamefont {Seipt},
		\citenamefont {Bulanov},\ and\ \citenamefont {Marklund}}]{Blackburn:2019lgk}%
	\BibitemOpen
	\bibfield  {author} {\bibinfo {author} {\bibfnamefont {T.~G.}\ \bibnamefont
			{Blackburn}}, \bibinfo {author} {\bibfnamefont {D.}~\bibnamefont {Seipt}},
		\bibinfo {author} {\bibfnamefont {S.~S.}\ \bibnamefont {Bulanov}}, \ and\
		\bibinfo {author} {\bibfnamefont {M.}~\bibnamefont {Marklund}},\ }\href@noop
	{} {\  (\bibinfo {year} {2019}{\natexlab{b}})},\ \Eprint
	{http://arxiv.org/abs/1904.07745} {arXiv:1904.07745 [physics.plasm-ph]}
	\BibitemShut {NoStop}%
	\bibitem [{Tac()}]{TackaSeipt}%
	\BibitemOpen
	\href@noop {} {}\bibinfo {note} {{D.~Seipt, private
			communication.}}\BibitemShut {Stop}%
	\bibitem [{\citenamefont {Kibble}\ \emph {et~al.}(1975)\citenamefont {Kibble},
		\citenamefont {Salam},\ and\ \citenamefont {Strathdee}}]{Kibble:1975vz}%
	\BibitemOpen
	\bibfield  {author} {\bibinfo {author} {\bibfnamefont {T.~W.~B.}\
			\bibnamefont {Kibble}}, \bibinfo {author} {\bibfnamefont {A.}~\bibnamefont
			{Salam}}, \ and\ \bibinfo {author} {\bibfnamefont {J.~A.}\ \bibnamefont
			{Strathdee}},\ }\href {\doibase 10.1016/0550-3213(75)90581-7} {\bibfield
		{journal} {\bibinfo  {journal} {Nucl. Phys.}\ }\textbf {\bibinfo {volume}
			{B96}},\ \bibinfo {pages} {255} (\bibinfo {year} {1975})}\BibitemShut
	{NoStop}%
	\bibitem [{\citenamefont {Harvey}\ \emph {et~al.}(2012)\citenamefont {Harvey},
		\citenamefont {Heinzl}, \citenamefont {Ilderton},\ and\ \citenamefont
		{Marklund}}]{Harvey:2012ie}%
	\BibitemOpen
	\bibfield  {author} {\bibinfo {author} {\bibfnamefont {C.}~\bibnamefont
			{Harvey}}, \bibinfo {author} {\bibfnamefont {T.}~\bibnamefont {Heinzl}},
		\bibinfo {author} {\bibfnamefont {A.}~\bibnamefont {Ilderton}}, \ and\
		\bibinfo {author} {\bibfnamefont {M.}~\bibnamefont {Marklund}},\ }\href
	{\doibase 10.1103/PhysRevLett.109.100402} {\bibfield  {journal} {\bibinfo
			{journal} {Phys. Rev. Lett.}\ }\textbf {\bibinfo {volume} {109}},\ \bibinfo
		{pages} {100402} (\bibinfo {year} {2012})},\ \Eprint
	{http://arxiv.org/abs/1203.6077} {arXiv:1203.6077 [hep-ph]} \BibitemShut
	{NoStop}%
	\bibitem [{\citenamefont {Nikishov}\ and\ \citenamefont
		{Ritus}(1964)}]{nikishov64}%
	\BibitemOpen
	\bibfield  {author} {\bibinfo {author} {\bibfnamefont {A.~I.}\ \bibnamefont
			{Nikishov}}\ and\ \bibinfo {author} {\bibfnamefont {V.~I.}\ \bibnamefont
			{Ritus}},\ }\href@noop {} {\bibfield  {journal} {\bibinfo  {journal} {Sov.
				Phys.--JETP}\ }\textbf {\bibinfo {volume} {19}},\ \bibinfo {pages} {529}
		(\bibinfo {year} {1964})}\BibitemShut {NoStop}%
	\bibitem [{\citenamefont {Brown}\ and\ \citenamefont
		{Kibble}(1964)}]{kibble64}%
	\BibitemOpen
	\bibfield  {author} {\bibinfo {author} {\bibfnamefont {L.~S.}\ \bibnamefont
			{Brown}}\ and\ \bibinfo {author} {\bibfnamefont {T.~W.~B.}\ \bibnamefont
			{Kibble}},\ }\href@noop {} {\bibfield  {journal} {\bibinfo  {journal} {Phys.
				Rev.}\ }\textbf {\bibinfo {volume} {133}},\ \bibinfo {pages} {A705} (\bibinfo
		{year} {1964})}\BibitemShut {NoStop}%
	\bibitem [{\citenamefont {Nikishov}\ and\ \citenamefont
		{Ritus}(1967)}]{nikishov67}%
	\BibitemOpen
	\bibfield  {author} {\bibinfo {author} {\bibfnamefont {A.~I.}\ \bibnamefont
			{Nikishov}}\ and\ \bibinfo {author} {\bibfnamefont {V.~I.}\ \bibnamefont
			{Ritus}},\ }\href@noop {} {\bibfield  {journal} {\bibinfo  {journal} {Sov.
				Phys.--JETP}\ }\textbf {\bibinfo {volume} {25}},\ \bibinfo {pages} {1135}
		(\bibinfo {year} {1967})}\BibitemShut {NoStop}%
	\bibitem [{\citenamefont {Narozhny\u{\i}}(1969)}]{narozhny69}%
	\BibitemOpen
	\bibfield  {author} {\bibinfo {author} {\bibfnamefont {N.~B.}\ \bibnamefont
			{Narozhny\u{\i}}},\ }\href@noop {} {\bibfield  {journal} {\bibinfo  {journal}
			{Sov. Phys.--JETP}\ }\textbf {\bibinfo {volume} {28}},\ \bibinfo {pages}
		{371} (\bibinfo {year} {1969})}\BibitemShut {NoStop}%
	\bibitem [{\citenamefont {Berestetskii}\ \emph {et~al.}(1982)\citenamefont
		{Berestetskii}, \citenamefont {Lifshitz},\ and\ \citenamefont
		{Pitaevskii}}]{landau4}%
	\BibitemOpen
	\bibfield  {author} {\bibinfo {author} {\bibfnamefont {V.~B.}\ \bibnamefont
			{Berestetskii}}, \bibinfo {author} {\bibfnamefont {E.~M.}\ \bibnamefont
			{Lifshitz}}, \ and\ \bibinfo {author} {\bibfnamefont {L.~P.}\ \bibnamefont
			{Pitaevskii}},\ }\href@noop {} {\emph {\bibinfo {title} {Quantum
				Electrodynamics (second edition)}}}\ (\bibinfo  {publisher}
	{Butterworth-Heinemann},\ \bibinfo {address} {Oxford},\ \bibinfo {year}
	{1982})\BibitemShut {NoStop}%
	\bibitem [{\citenamefont {Baier}\ and\ \citenamefont
		{Katkov}(1968)}]{BaierA0B0}%
	\BibitemOpen
	\bibfield  {author} {\bibinfo {author} {\bibfnamefont {V.}~\bibnamefont
			{Baier}}\ and\ \bibinfo {author} {\bibfnamefont {V.}~\bibnamefont {Katkov}},\
	}\href@noop {} {\bibfield  {journal} {\bibinfo  {journal} {Sov.Phys.JETP}\
		}\textbf {\bibinfo {volume} {26}},\ \bibinfo {pages} {1238} (\bibinfo {year}
		{1968})}\BibitemShut {NoStop}%
	\bibitem [{\citenamefont {Khokonov}\ and\ \citenamefont
		{Khokonov}(2005)}]{Khok}%
	\BibitemOpen
	\bibfield  {author} {\bibinfo {author} {\bibfnamefont {A.~K.}\ \bibnamefont
			{Khokonov}}\ and\ \bibinfo {author} {\bibfnamefont {M.}~\bibnamefont
			{Khokonov}},\ }\href@noop {} {\bibfield  {journal} {\bibinfo  {journal}
			{Tech.Phys.Lett.}\ }\textbf {\bibinfo {volume} {31}},\ \bibinfo {pages} {154}
		(\bibinfo {year} {2005})}\BibitemShut {NoStop}%
	\bibitem [{\citenamefont {Dinu}\ \emph {et~al.}(2016)\citenamefont {Dinu},
		\citenamefont {Harvey}, \citenamefont {Ilderton}, \citenamefont {Marklund},\
		and\ \citenamefont {Torgrimsson}}]{Dinu:2015aci}%
	\BibitemOpen
	\bibfield  {author} {\bibinfo {author} {\bibfnamefont {V.}~\bibnamefont
			{Dinu}}, \bibinfo {author} {\bibfnamefont {C.}~\bibnamefont {Harvey}},
		\bibinfo {author} {\bibfnamefont {A.}~\bibnamefont {Ilderton}}, \bibinfo
		{author} {\bibfnamefont {M.}~\bibnamefont {Marklund}}, \ and\ \bibinfo
		{author} {\bibfnamefont {G.}~\bibnamefont {Torgrimsson}},\ }\href {\doibase
		10.1103/PhysRevLett.116.044801} {\bibfield  {journal} {\bibinfo  {journal}
			{Phys. Rev. Lett.}\ }\textbf {\bibinfo {volume} {116}},\ \bibinfo {pages}
		{044801} (\bibinfo {year} {2016})},\ \Eprint
	{http://arxiv.org/abs/1512.04096} {arXiv:1512.04096 [hep-ph]} \BibitemShut
	{NoStop}%
\end{thebibliography}
\end{document}